\documentclass[aps,floatfix,showpacs,showkeys,amsmath]{revtex4}
\usepackage{color,graphicx}% Include figure files
\usepackage{dcolumn}% Align table columns on decimal point
\usepackage{bm}% bold math
\usepackage{epsfig}
\usepackage{supertabular}
\usepackage[bookmarksopen]{hyperref}
\usepackage{url}
\usepackage{graphicx}
\usepackage{float}
\usepackage[caption = false]{subfig}
\usepackage{verbatim}

\def\be{\begin{equation}}
\def\ee{\end{equation}}
\def\nn {\nonumber}
\def\bea {\begin{eqnarray}}
\def\eea {\end{eqnarray}}
\def\pnn{\pi\scriptscriptstyle{N N}}
\def\gg{\gamma}
\def\g5{\gamma^5}
\def\slkp{k_{||}\!\!\!\!\!\!/ ~}
\def\slkpr{k_{\perp}\!\!\!\!\!\!\!/ ~~}

\def\slpp{p_{||}\hspace{-0.35cm}/~}

\def\slq{q\hspace{-0.16cm}/}
\def\slk{k\hspace{-0.16cm}/}
\def\qp{q_{||}}
\def\kp{k_{||}}
\def\pp{p_{||}}
\def\qpr{q_{\perp}}
\def\kpr{k_{\perp}}
\def\a{\alpha}
\def\D{\Delta}
\def\d{\delta}
\usepackage{axodraw4j}
\usepackage{pstricks}
\usepackage{color}
\begin{document} 
 \title{\Large {\bf Mass modification of hot pions in magnetized dense medium }}
\renewcommand{\thefootnote}{\alph{footnote}}
\bigskip
\bigskip
\author{Arghya Mukherjee$^{a,c}$}
\email{arghya.mukherjee@saha.ac.in}
\author{Snigdha Ghosh$^{b,c}$}
\email{snigdha.physics@gmail.com}
\author{Mahatsab Mandal$^{a,d}$}
\email{mahatsab@gmail.com}
\author{ Pradip Roy$^{a,c}$}
\email{pradipk.roy@saha.ac.in}
\author{Sourav Sarkar$^{b,c}~$}
\email{sourav@vecc.gov.in}

\affiliation{$^a$Saha Institute of Nuclear Physics, 1/AF Bidhannagar
Kolkata - 700064, India}
\affiliation{$^b$Variable Energy Cyclotron Centre
1/AF Bidhannagar, Kolkata 700 064,
India}
\affiliation{$^c$Homi Bhabha National Institute, Training School Complex, Anushaktinagar, Mumbai - 400085, India}
\affiliation{$^d$Government General Degree College at Kalna-I, Burdwan, West Bengal - 713405, India}

\begin{abstract}
A phenomenological pion-nucleon interaction is used to obtain pionic  mass modification in presence of constant homogeneous magnetic field background at  
finite temperature and chemical potential in the real time formalism of thermal field theory. The  
magnetically modified propagator in its complete form is used to obtain the one loop self-energy for pions. 
%%%%%%%%%%%%%%%%%%%%%%%%%%%%%%%%%%%%%%%%%%%%%%%%%%%%%%%%%%%%%%%%%%%%%%%%%%%%%%%%%%%%%%%%%%%%%%%%%%%%%%%%%%%%%%%%%%%%%%%%%
For charged pions we find that the effective mass increases with the magnetic field at given temperature and chemical potential. Since the transverse momentum
of charged pion is quantized and its contribution to Dyson-Schwinger Equation is large compared to the loop correction, the charged pion mass remains constant with
both temperature and chemical potential for a given landau level. In order to unveil the role of the real part of the self-energy, we also calculate the effective mass neglecting the 
trivial shift. %%%%%%%%%%%%%%%%%%%%%%%%%%%%%%%%%%%%%%%%%%%%%%%%%%%%%%%%%%%%%%%%%%%%%%%%%%%%%%%%%%%%%%%%%%%%%%%%%%%%%%%%%%%%%%%%%%%%%
The effective mass for charged pions 
shows an oscillatory behaviour which is attributed to the thermal contribution of the self-energy. It is  argued that the magnetic field dependent vacuum 
contribution to the self-energy influences the  behaviour of the effective mass both qualitatively and quantitatively. We also find that very large field is 
necessary for neutral pions to condense.
\end{abstract}
\maketitle

\section{Introduction}
Understanding quantum chromodynamics(QCD) in presence of magnetic background has gained a lot of contemporary research interest \cite{ln871}. It is important to study QCD in presence of 
external magnetic field not only  for  its  relevance to astrophysical phenomena \cite{prep442,prl95,prd76,prl100,prdN76,prl105,prd82} but also due to  the possibility 
of strong magnetic field production in 
non-central heavy-ion collisions \cite{ijmp} which sets the stage for investigation of these  magnetic modifications. Although the background fields  produced in RHIC and 
LHC are 
much smaller in comparison with  the  field strengths that prevailed during the cosmological electro-weak phase transition which   may reach up to $eB\approx200m_\pi^2$
\cite{plb265}, 
they are strong enough to cast significant influence on the hadronic properties which bear the information of the chiral phase transition.  Moreover, among the hadrons, 
mesons possess  direct connection to the chiral phase transition\cite{prdW86} that make them more important candidates than baryons for the  understanding of the phenomena. 
 At vanishing chemical potential, modification due to the presence of magnetic background
can be obtained from first principle using lattice QCD simulations \cite{prd82M,prd83} which shows monotonic increase in critical temperature with 
the increasing magnetic field, also 
known as magnetic catalysis(MC). The effects of external magnetic field on the chiral phase transition  has been studied using different effective models in recent years 
\cite{tmph90,prd81,prd84,prc83,prd83M,prc83A,prd82R,prd83K,prd85,jhep08,prd82A,prd85S,prd84D,prd86E}.
QCD being a confining theory at low energies, effective theories are employed to describe the low energy behaviour of the strong interaction. 
These effective field theoretic models in general contain a few parameters which can be  fixed from experimental inputs. Although most of the model calculations are 
in support of 
MC,  different lattice results had shown inverse magnetic catalysis(IMC) where critical temperature follows
the opposite trend \cite{jhep1202,prdB86,jhep04,prd90,prc90,prd91,plb759}. It was  pointed out in \cite{jhep1304} that IMC is  attributed to the dominance of the sea 
contribution over the valence  contribution of the quark condensate.  
The sea  effect has not been incorporated even in the  Polyakov loop extended versions of  Nambu--Jona-Lasinio (PNJL) model 
and Quark-Meson(PQM) model which might be a possible reason of  the disagreement. To investigate the apparent contradiction, significant works have been done \cite{rmp88} in quest of proper modifications of the effective models, most of which are
focused on the magnetic field dependency of the coupling constants or other
 magnetic field dependent parameters in the model.  Very recently, 
IMC has been observed in   NJL model,
with Pauli-Villars regularization scheme \cite{plb758} which gives markedly different behaviour in comparison with the usual  soft-cutoff approach.

It has been argued in \cite{prd79} that the study of pion-nucleon interaction
plays an important role in the behavioural description of the deconfinement
critical temperature in terms of pion mass and isospin. Pion mass modification in presence of 
magnetic background has  been calculated  employing   chiral perturbation theory in Ref.\cite{prd86}. 
 It has been shown that in presence of magnetic field, the charged pions are no longer the Goldstone modes and the critical temperature of chiral phase 
 transition shows magnetic catalysis. In NJL approach  \cite{epjc76}, it  has been found that there exists a sudden leap in the effective masses of 
 charged pions near the same critical temperature from where the $\sigma$ and $\pi^0$ meson become nearly degenerate. Here also, the pseudo-critical temperature is found to 
 increase with the increasing magnetic field.
%mass degeneracy between $\sigma$ and $\pi$  is  not precise as it was in case of the zero magnetic field. 
In Ref.\cite{colucci} pion effective mass variation with $eB$ has been observed for large magnetic fields where
Lowest Landau Level(LLL) approximation is reasonable. Also in Ref.\cite{aguirre}, the medium modification of pion effective mass has been obtained in a self-consistent 
way with LLL approximation.  
However, it is a common trend to ignore the magnetic field dependent {\it vacuum} contribution of the self-energy function in case of mass and dispersion calculations.
It has been shown \cite{spa,me} (also
see the references therein) that
it can have significant influence on  mesonic properties like effective mass, dispersion relation, decay width and spectral function.
Merely on grounds of simplicity, it appears unreasonable  to neglect this $eB$ dependent vacuum contribution apriori unless one compares the dependencies on other 
external parameters with it. It is also interesting to observe the interplay between the medium  and the vacuum effect of the external magnetic field  to 
find the complete $eB$ dependence of the physical
properties.  
In this article we revisit the 
 mass modification of pions in presence of finite temperature and  chemical potential  in a homogeneous magnetic background with  
a well known pion-nucleon interaction in isospin-symmetric nuclear matter. Unlike NJL model, here pionic fields are treated 
as elementary. However, the non-trivial  mass correction  in presence of magnetic field occurs  due to  the modification of the nucleon propagators. The influence of the 
magnetic field dependent vacuum contribution in case of the pseudo-vector pion-nucleon interaction  has been studied in detail.  
In the Dyson-Schwinger formalism, instead of restricting ourselves to the strong/weak field region, 
the full 
propagator is used to obtain  the pion self-energy. %%%%%%%%%%%
Since the charged pion transverse momentum is quantized in magnetic field, we obtain Landau Level(LL) dependent self-energy and the corresponding Dyson-Schwinger 
Equation (DSE)  
is modified because of the presence of the transverse momentum. Thus we obtain LL dependent effective mass for charged pions. To see the importance of $eB$ dependent vacuum 
contribution we neglect the trivial shift. In case of neutral pions, although the evaluation is restricted to the special case where the external pion momentum is parallel
to the field 
direction, this restriction does not put any constraint in case of mass calculation. More specifically, pion effective mass has been obtained with full magnetic field 
dependence up to one loop order. Effective mass variation with external magnetic field due to $eB$ dependent vacuum is compared with the $eB$ dependent thermal 
contribution. It is argued that neglecting vacuum contribution may even influence the qualitative predictions of the effective mass dependences for different pion species. 

The article is organized as follows. In Sec.II we discuss the formalism for calculating the one loop pion self-energy function for phenomenological 
pion-nucleon interaction  in presence of constant external magnetic field  in
dense thermal medium.  The section comprises  two subsections, one for the charged pions   and the other for the  neutral pion  where the magnetic field dependent 
 vacuum contribution and thermal 
contribution   of the self-energy  of the corresponding species  are calculated.  The Dyson-Schwinger  equation that relates the
effective mass with the real part of the self-energy is also obtained. Pion mass variation with respect to the independent parameters are presented in Sec.III. 
The effect of incorporation of magnetic field dependent vacuum part is also discussed. Finally  we summarize our work in Sec.IV.   
\section{Formalism}
\begin{center}
\begin{figure}
 {\includegraphics[width=6.5in,angle=0]{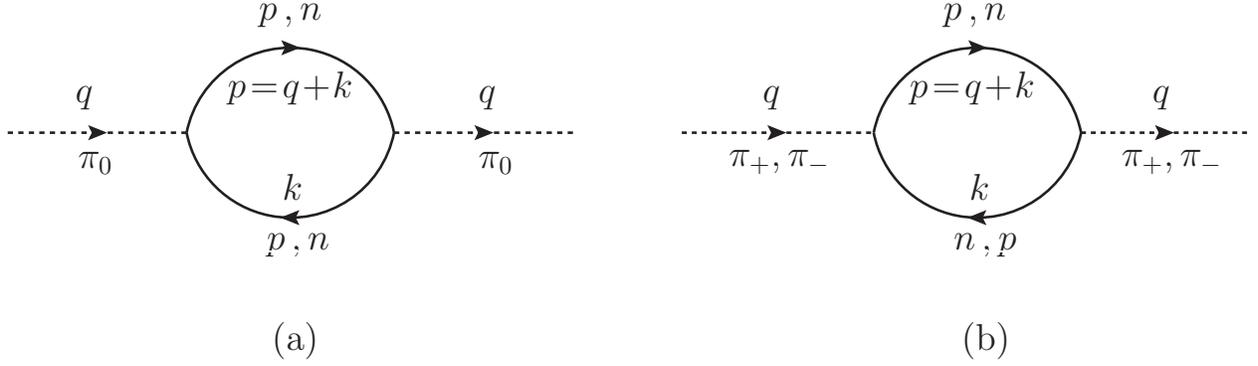}}
  \caption{One loop Feynman diagrams for pion-nucleon interaction.}
\label{feyn}
\end{figure}
\end{center}

The Dyson-Schwinger Equation(DSE) for the effective propagator of pion is given by 
\bea
D^{-1}(q)&=&D_0^{-1}(q)-\Pi(q)
\eea
where $D_0^{-1}(q)=q^2-m_\pi^2+i\epsilon$  and $\Pi(q)$ is the pion self-energy. One can obtain  the   effective mass by finding the pole of $D(q)$. Here, we are 
interested in finding the thermal modification of  pion mass in presence of constant external  magnetic field along with finite baryon 
density due to the effective pion-nucleon
interaction, given by~\cite{prc62}  
\bea
\mathcal{L}^{int} _{\pnn}&=&-\frac{f_{\pnn}}{m_\pi}\bar{\psi}\g5\gg^\mu(\vec{\tau}\cdot\partial_\mu\vec{\pi})\psi.
\eea
where $\psi$ is the two component nucleon field and $\vec{\tau}=(\sigma_x,\sigma_y,\sigma_z)$,  with $\sigma_a$  denoting the $a$\,th Pauli spin matrix. 
The pionic fields are represented by the isovector  $\vec{\pi}$. Expanding the interaction Lagrangian, 
one finds the  Feynman diagrams for the one loop self-energy of pions as given in Fig.\ref{feyn}. 

In the real time formalism of thermal field theory, the propagators as well as the one loop  self-energy function assume a $2\times2$ matrix structure. The 11-components  of the matrices 
for neutral and charged pions are given by  

\bea
\Pi^{11}_0(q,\mu_N,T)&=&-ig^2_{\pnn}\int\frac{d^4k}{(2\pi)^4}\mbox{Tr}\Big[\g5\slq S^{11}_p(k)\g5\slq S^{11}_p(k+q)
\Big]-ig^2_{\pnn}\int\frac{d^4k}{(2\pi)^4}\mbox{Tr}\Big[\g5\slq S^{11}_n(k)\g5\slq S^{11}_n(k+q)
\Big]\nn\\
\Pi^{11}_+(q,\mu_N,T)&=&-2ig^2_{\pnn}\int\frac{d^4k}{(2\pi)^4}\mbox{Tr}\Big[\g5\slq S^{11}_n(k)\g5\slq S^{11}_p(k+q)
\Big]\nn\\
\Pi^{11}_-(q,\mu_N,T)&=&-2ig^2_{\pnn}\int\frac{d^4k}{(2\pi)^4}\mbox{Tr}\Big[\g5\slq S^{11}_p(k)\g5\slq S^{11}_n(k+q)
\Big]
\label{mother}
\eea

where $S^{11}_p(k)$ and  $S^{11}_n(k)$ are the 11-components of the  thermal propagators for proton and neutron respectively. 
The real time thermal propagators can be decomposed into two parts as \cite{olivo}
\bea
S^{11}_p(k)&=&S_p(k)-2i\eta(k)~\mbox{Im} S_p(k)\nn\\
\hspace{-0.5cm}\mbox{and}\hspace{0.5cm}S^{11}_n(k)&=&S_n(k)-2i\eta(k)~\mbox{Im} S_n(k)\nn\\
\hspace{-0.8cm}\mbox{with}\hspace{0.8cm}\eta(k)&=&\theta(k^0)n_k^++\theta(-k^0)n_k^-\nn\\
\hspace{-1cm}\mbox{and}\hspace{1cm}n^{\pm}_k&=&\frac{1}{e^{\beta(\omega _k\mp\mu_N)}+1}.
\eea
Here, $S_p(k)$ is the momentum space representation of the fermionic propagator in presence of magnetic field, $\theta$ denoting the unit step function and 
$\beta=\frac{1}{T}$ is the inverse temperature in natural unit. Fermionic propagators in presence of magnetic field possess 
a phase factor which  can not be taken as translationally invariant in general. However, in the current context, the phase factor can be removed by suitable gauge 
transformation \cite{prd92} and we can work with the momentum space representation of the translationally invariant part which  is given by \cite{nuphyb462}
\bea
S_p(k)&=&-\sum_{n=0}^{n=\infty}(-1)^n e^{-\alpha}\frac{\big[(\slkp+m)\{(1-i\gg^1\gg^2)L_n(2\alpha)
-(1+i\gg^1\gg^2)L_{n-1}(2\alpha)\}-4\slkpr L^1_{n-1}(2\alpha)\big]}{\slkp^2-m^2-2n eB+i\epsilon}
\eea
where $\a=-\kpr^2/eB$ and  $L_n\equiv L_n^0$ with $L_n^\a$ representing the generalized Laguerre polynomials. The $\epsilon$ in the denominator 
is an infinitesimal positive parameter. It should be 
mentioned here that in this article we use $g^{\mu\nu}_{||}={\rm diag}(1,0,0,-1)$ 
and $g^{\mu\nu}_{\perp}={\rm diag}(0,-1,-1,0)$ with metric defined as $g^{\mu\nu}=g^{\mu\nu}_{||}+g^{\mu\nu}_{\perp}$. 
A general four vector can be decomposed as 
$a^\mu=a^\mu_{||}+a^\mu_{\perp}$ with $a_{||}^2=a_0^2-a_3^2$ and $a_{\perp}^2=-a_1^2-a_2^2$.
The imaginary part of the propagator is
\bea
\mbox{Im} S_p(k)&=&\pi\sum_{n=0}^{n=\infty}
(-1)^n e^{-\alpha}
\big[(\slkp+m)\{(1-i\gg^1\gg^2)L_n(2\alpha)-(1+i\gg^1\gg^2)L_{n-1}(2\alpha)\}-4\slkpr
 L^1_{n-1}(2\alpha)\big]\nn\\
 &\times &\delta(\slkp^2-m^2-2n eB).
 \eea
However, the  neutron propagator $S_n(k)$ is not influenced by  the presence of the magnetic field and is given by  
 \bea
 S_n(k)&=&-\frac{\slk+m}{k^2-m^2+i\epsilon}\,\,\,\,\mbox{and}\nn\\
 \mbox{Im} S_n(k)&=&\pi(\slk+m)\delta(k^2-m^2).
\eea
Now,  the propagators have two distinct parts, one with the thermal distribution function  and another without it. On this basis the self-energy function can be 
expressed as a sum of three different portions given by 
\bea
\Pi^{11}_{0,\pm}&=&(\Pi_{0,\pm})_{\mbox{vac}}+(\Pi^{}_{0,\pm})_\eta +(\Pi^{}_{0,\pm})_{\eta^2}.
\eea
The term with quadratic dependence on the distribution function is purely imaginary. 
As we are only interested in the real part of the self-energy,  we have 
\bea
\mathcal{R}\mbox{e}(\Pi^{11}_{0,\pm})\equiv\mathcal{R}\mbox{e}(\overline{\Pi}_{0,\pm})
&=&\mathcal{R}\mbox{e}(\Pi^{}_{0,\pm})_{\mbox{vac}}+\mathcal{R}\mbox{e}(\Pi^{}_{0,\pm})_\eta
\eea
where $\overline{\Pi}$ represents the 11-component of the diagonal self-energy matrix (see e.g \cite{sirbook}). 
Let us now consider the explicit forms of the real part of the 
self-energy for  charged and neutral mesons separately.
%In both of the cases, it is straight forward to obtain the thermal contributions and only final results are shown
%whereas necessary details are provided for the vacuum contributions.

%%%%%%%%%%%%%%%%%%%%%%%%%%%%%%%%%%%%%%%%%%%%%%%%%%%%%%%%%%%%%%%%%%%%%%%%%%%%%%%%%%%%%%%%%%%%%%%%%%%%%%%%%%%%%%%%%%%%%%%%%%%%%%%%%%%%%%
\subsection{Charged pions}
 The medium independent vacuum  self-energy will be same for the  charged pions $\pi^+$ and $\pi^-$ and can be obtained as follows.
\bea
(\Pi^{}_{+})_{\mbox{vac}}&=&-2ig^2_{\pnn}\int\frac{d^4 k}{(2\pi)^4}\mbox{Tr}\Big[\g5\slq S_{n}(k)\g5\slq S_{p}(p=q+k)\Big]\nn\\
&=&-2ig^2_{\pnn}\int\frac{d^2\kp}{(2\pi)^2}\frac{d^2\kpr}{(2\pi)^2}\sum_{n=0}^\infty\frac{(-1)^ne^{-\alpha _p}\mbox{Tr}
\Big[\g5\slq(\slk+m)\g5\slq D_n(q+k)\Big]}{(k^2-m^2+i\epsilon)(\slpp^2-m^2-2neB+i\epsilon)}\nn\\
&=&i\sum_{n=0}^\infty\int\frac{d^2\kpr}{(2\pi)^2}\mbox{I}(n,\kpr ^2,\qp ^2,\qpr^2)\hspace{1cm}\mbox{where}\label{pi+}\\
\mbox{I}(n,\kpr ^2,\qp ^2,\qpr^2)&=&\int\frac{d^2\kp}{(2\pi)^2}\frac{\mathcal{N}_n}{(k^2-m^2+i\epsilon)(\pp^2-m^2-2neB+i\epsilon)}\hspace{1cm}
\,\,\,\mbox{with}\nn\\
\mathcal{N}_n&=&-2g_{\pnn}^2(-1)^ne^{-\a_p}\mbox{Tr}
\Big[\g5\slq(\slk+m)\g5\slq D_n(q+k)\Big]\nn\\
&=&-8g_{\pnn}^2(-1)^ne^{-\a_p}\Big[4L^1_{n-1}(2\a_p)\{(\qp^2-\qpr^2-2\qpr\!\!\cdot\kpr)(\qpr\!\cdot\kpr) +q^2\kpr^2-2(\qpr^2+\qpr\!\!\cdot\kpr)(\qp\cdot\kp)\}\nn\\
&+&\big(L_n(2\a_p)-L_{n-1}(2\a_p)\big)\{-m^2q^2+2\qp^2(\qpr\!\!\cdot\kpr)-q^2\kp^2+2(\qp\cdot\kp)^2\nn\\
&+&(\qp^2-\qpr^2+2\qpr\!\!\cdot\kpr)(\qp\cdot\kp)\}\Big].
\label{i}
\eea 
Here $\a _p=-p_{\perp}^2/eB$ and $p=q+k$ with $q$ being the external momentum of pions.
 The 2-dimensional $\kp$ integration can be performed using standard Feynman parametrization and dimensional regularization technique to obtain
 \bea
 I(n,\kpr,\qp^2,\qpr^2)&=&-8g_{\pnn}^2(-1)^n e^{-\a_p}\frac{i}{4\pi}\int_0^1dx\Big[4L_{n-1}^1(2\a_p)\frac{A}{\D_n}+\big(L_n(2\a_p)-L_{n-1}(2\a_p)\big)\Big(\frac{B_n}{\D_n}
 -\qpr^2\ln\frac{\D_n}{\mu_0}\Big)\Big]\nn\\
 \mbox{with}\hspace{1cm}A&=&q^2\kpr^2+2x\qp^2\qpr^2+\{(1+2x)\qp^2-\qpr^2-2(\qpr\!\cdot\kpr)\}(\qpr\!\cdot\kpr)\nn\\
 B_n&=&-2 m^2 q^2-x(1-x)(\qp^4-\qp^2\qpr^2-\qp^2 q^2)+2(1-x)\qp^2(\qpr\!\cdot\kpr)+(1-x)q^2\kpr^2-2 n x eB q^2\nn\\
 &=&B_0-2 n x eB q^2\nn\\
 \D_{n}&=&m^2-x(1-x)\qp^2-i\epsilon-(1-x)\kpr^2+2 n x eB\nn\\
 &=&\D-(1-x)\kpr^2+2n x eB\label{ifinal}.
 \eea
where $\mu_0$ is the scale which appears in the process of dimensional regularization. Now, with this $I(n,\kpr,\qp^2,\qpr^2)$,  the summation in  Eq.(\ref{pi+}) 
can be taken inside the $\kpr$ integral which gives  
\bea
(\Pi_+)_{\mbox{vac}}&=&\frac{2}{\pi}g^2_{\pnn}\int_0^1dx\int\frac{d^2\kpr}{(2\pi)^2}e^{-\a_p}\Big[4A\mathcal{S}_1+\mathcal{S}_2-\qpr^2\mathcal{S}_3\Big]
\hspace{1cm}\mbox{where}\nn\\
\mathcal{S}_1&=&\sum_n^\infty(-1)^nL_{n-1}^1(2\a_p)\frac{1}{\D_n}\nn\\
\mathcal{S}_2&=&\sum_n^\infty(-1)^n\big(L_n(2\a_p)-L_{n-1}(2\a_p)\big)\frac{B_n}{\D_n}\nn\\
\mathcal{S}_3&=&\sum_n^\infty(-1)^n\big(L_n(2\a_p)-L_{n-1}(2\a_p)\big)\ln\frac{\D_n}{\mu_0}\label{sum}.
\eea
It is possible to find compact expressions for these summations by  casting  them  into known series of Laguerre polynomials as discussed in detail in the appendix.
The final expression of the vacuum contribution of the charged pion self-energy  is given by 
\bea
(\Pi_+)_{\mbox{vac}}&=&\frac{g^2_{\pnn}}{2\pi^2}\int_0^1\int_0^1dx\,\frac{dz}{\mu}z^{\frac{\D}{\mu}-y(1-x)\frac{\qpr^2}{\mu}-1}\frac{1}{\eta}\Big[\mbox{sech}^2
(x\frac{eB}{\mu}\ln z)\{\frac{1}{\eta}(\qp^2-xq^2)+\qp^2\qpr^2(y^2+x-y-xy^2)\nn\\
&+&\qpr^4(y-x-y^2+2xy-xy^2)\}+2m^2q^2+2\qp^2\qpr^2(1-x)(y-x)-(1-x)y^2q^2
\qpr^2\nn\\
&-&(1-x)\frac{q^2}{\eta}+\frac{\mu}{\ln z}\qpr^2\Big]\hspace{1cm} \mbox{where}\nn\\
\eta&=&(1-x)\frac{\ln z}{\mu}+\frac{1}{eB}\mbox{tanh}(x\frac{eB}{\mu}\ln z)\nn\\
y&=&\frac{1}{\eta eB}\mbox{tanh}(x\frac{eB}{\mu}\ln z)\label{finalpi+}.
\eea
One can observe here that instead of a 2-dimensional $\kpr$ integral and an infinite series summation, now we have one integration over the parameter $z$ which is  more 
convenient for numerical evaluation. Another important feature is that the expression is not in the form of any polynomial of $eB$ which signifies its non-perturbative
character. It should be pointed out that the scale $\mu$ present here appears in the process of  parametrization with $z$ and is different from the scale $\mu_0$ that
appeared from dimensional regularization of $\kp$. It can be shown from Eq.(\ref{finalpi+}) that at  $eB=0$ the self energy becomes 
\bea
(\Pi_+)_{\mbox{vac}}(eB=0)&=&g^2_{\pnn}\frac{m^2q^2}{\pi^2}\int_0^1 dx\, \ln \Big(\frac{\D}{\mu}\Big)\hspace{0.5cm}
\eea
which is exactly twice the contribution of $nn$ loop in case of neutral pions as will be seen later. Thus the $(\Pi_+)_{\mbox{vac}}$ can be decomposed as
\bea
(\Pi_+)_{\mbox{vac}}&=&(\Pi_+)_{\mbox{vac}}(eB\ne0)+(\Pi_+)_{\mbox{vac}}(eB=0)
\eea
where $(\Pi_+)_{\mbox{vac}}(eB\ne0)$ represents the external $eB$ dependent part of the self-energy and will be used in the DSE to obtain the effective mass.
The procedure to obtain the thermal part of the self-energy is relatively simpler and only the final expressions are presented.
The real part of the thermal contribution of the self-energy for $\pi^+$   is given by
\bea
\mathcal{R}\mbox{e}(\Pi_+)_\eta&=&\sum_{l=0}^{\infty}\int\frac{d^3k}{(2\pi)^3}
\mathcal{P}\Big[\frac{\mathcal{N}_{l}(k^0=-q^0+\omega^l_p)n^{+}_{p,l}}{2\omega^l_p\{(q^0-\omega^l_p)^2-(\omega_k)^2\}}
+\frac{\mathcal{N}_{l}(k^0=-q^0-\omega^l_p)n^{-}_{p,l}}{2\omega^l_p\{(q^0+\omega^l_p)^2-(\omega_k)^2\}}\nn\\
&+&\frac{\mathcal{N}_{l}(k^0=\omega_k)n^{+}_{k}}{2\omega_k\{(q^0+\omega_k)^2-(\omega^l_p)^2\}}+
\frac{\mathcal{N}_{l}(k^0=-\omega_k)n^{-}_{k}}{2\omega_k\{(q^0-\omega_k)^2-(\omega^l_p)^2\}}\Big]
\label{nl}
\eea
where the expression for $\mathcal{N}_l$ is given in Eq.(\ref{i}) for the dummy index $l=n$.
One can observe from Eq.(\ref{mother}) that 
\bea
\Pi^{11}_+(-q,\mu_N,T)&=&\Pi^{11}_-(q,\mu_N,T).
\label{pi+pi-}
\eea
Thus the expression of $\mathcal{R}\mbox{e}(\Pi_-)_\eta$  can be easily obtained from $\mathcal{R}\mbox{e}(\Pi_+)_\eta$ by successively 
changing $k\rightarrow k-q$ and $q\rightarrow-q$. It should be mentioned here that unlike the vacuum case, the infinite sum could not be performed analytically and as a 
result the Laguerre polynomials remain in the numerator within the single sum structure of the thermal contribution.

%%%%%%%%%%%%%%%%%%%%%%%%%%%%%%%%%%%%%%%%%%%%%%%%%%%%%%%%%%%%%%%%%%%%%%%%%%%%%%%%%%%%%%%%%%%%%%%%%%%%%%%%%%%%%%%%%%%%%%%%%%%%%%%%%%%%%%%%%%%%%%%%%%%%

As in this case the external particles are charged, the external transverse momentum  suffers Landau quantization in presence of $eB$. 
Thus to obtain the effective mass  of $\pi^{\pm}$ as a function of temperature, chemical potential and external magnetic field,
we now solve the DSE of charged pions  given by 
\bea
m_{\pi^{\pm}}^{*\,2}-m_{\pi^{\pm}}^{\,2}+\mathcal{R}\mbox{e}\Pi_{\pm}(m_{\pi^{\pm}}^{*\,2},\qpr^2=(2n+1)eB,eB)-(2n+1)eB=0
\label{dsepm}
\eea
where $\mathcal{R}\mbox{e}\Pi_{\pm}$ contains the sum of explicit $eB$ dependent vacuum and thermal contributions. 
Here, $m_{\pi^{\pm}}^{}$ and  $m_{\pi^{\pm}}^{*}$ denotes  the renormalized  mass  and the effective mass of the charged pions respectively.

\subsection{Neutral pions}

At first  we consider  the magnetic field dependent vacuum contributions from $pp$ and $nn$ loops. 
 Here, the vacuum contribution  refers to the part of the self-energy which is independent of the thermal 
distribution functions and can be written as 
\bea
(\Pi^{}_0)_{\mbox{vac}}&=&(\Pi^{}_0)^{pp}_{\mbox{vac}}+(\Pi^{}_0)_{\mbox{vac}}^{nn}.
\eea
For the  $pp$ loop 
\bea 
(\Pi^{}_0)^{pp}_{\mbox{vac}}&=&-ig^2_{\pnn}\int\frac{d^4k}{(2\pi)^4}\mbox{Tr}\Big[\g5\slq S_p(k)\g5\slq S_p(p=q+k)\Big]\nn\\
&=&-ig^2_{\pnn}\int\frac{d^2\kp}{(2\pi)^2}\frac{d^2\kpr}{(2\pi)^2}\sum_{n,l=0}^{\infty}\frac{(-1)^{n+l}e^{-({\a}_k+\a_p)}\mbox{Tr}\Big[\g5\slq D_n(k)\g5\slq D_l(p)\Big]}{(\kp^2-m^2-2neB+i\epsilon)(\pp^2-m^2-2leB+i\epsilon)}\nn\\
\mbox{where}\hspace{1cm}D_n(k)&=&\big[(\slkp+m)\{(1-i\gg^1\gg^2)L_n(2\alpha)-(1+i\gg^1\gg^2)L_{n-1}(2\alpha)\}-4\slkpr L^1_{n-1}(2\alpha)\big].
\eea

At this point, one can observe that with $q_x=q_y=0$ the $\kpr$ integration can be done analytically and standard Feynman parametrization
technique can be applied to obtain  
\bea
(\Pi^{}_0)^{pp}_{\mbox{vac}}&=&-i\sum_{n,l=0}^{\infty}\int\frac{d^2\kp}{(2\pi)^4}\int_0^1 dx\frac{\mathcal{N}^{n,l}_{||}(\kp)}{\Big[(\kp+x\qp)^2-\D_{nl}\Big]^2}.
\eea
Dropping the odd terms after  the momentum shift $\kp\rightarrow\kp-x\qp$, we get
\bea
\mathcal{N}^{n,l}_{||}(\kp)&=&4\pi g^2_{\pnn}eB\Big[4eB\qp^2n\delta_{n-1,l-1}+(\delta_{n,l}+\delta_{n-1,l-1})\nn\\&&\times[\qp^2(-m^2+x^2\qp^2-x\qp^2)+2(\kp\cdot\qp)^2-\qp^2\kp^2]\Big]\nn\\
\mbox{with}\hspace{1cm}\D_{nl}&=&m_n^2-x(1-x)\qp^2+x(m_l^2-m_n^2)\nn\\
\mbox{where}\hspace{1cm}m_n^2&=&m^2+2neB-i\epsilon.
%\{2(\qp\cdot\kp)^2+\qp^2(-\kp^2-m^2+\qp\cdot\kp)\}\Big]
\eea
After the momentum integration we obtain 
\bea
(\Pi^{}_0)^{pp}_{\mbox{vac}}&=&\frac{eB}{4\pi^2}g^2_{\pnn}\qp^2\int_0^1dx\sum_{n,l=0}^\infty\frac{\Big[4eBn\d_{n-1,l-1}+(\d_{n,l}+\d_{n-1,l-1})\{-m^2-m_n^2-x(m_l^2-m_n^2)\}\Big]}{\D_{nl}}\nn\\
&=&\sum_{n,l=0}^\infty\frac{\Big[4eBn\d_{n-1,l-1}+(\d_{n,l}+\d_{n-1,l-1})\{-m^2-m_n^2-x(m_l^2-m_n^2)\}\Big]}{\D_{nl}}\nn\\
&=&\sum_{n,l=1}^\infty\frac{\Big[4eBn\d_{n,l}+2\d_{n,l}\{-m^2-m_n^2-x(m_l^2-m_n^2)\}\Big]}{\D_{nl}}
-\frac{2m^2}{\D}\nn\\
&&\hspace{0.5cm}\mbox{with}\hspace{0.5cm}\D_{00}=\D =m^2-x(1-x)\qp^2-i\epsilon\nn\\
&=&-\frac{2m^2}{eB}\sum_{n=1}^\infty\frac{1}{n+\frac{\D}{2eB}}-\frac{2m^2}{\D}\nn\\
&=&-\frac{2m^2}{eB}\big[\frac{1}{\epsilon}-\gamma-\psi(\frac{\D}{2eB})-\frac{2eB}{\D}\Big]-\frac{2m^2}{\D}
\hspace{0.3cm}\mbox{where}\hspace{0.5cm}\psi(\alpha) =\frac{\mbox{d} \ln(\Gamma (\alpha))}{\mbox{d} \alpha}.
\eea

It should be noted  that $\d_{n,-1}=0$ here as Laguerre polynomial with negative index is taken to be zero. 
%The function $\mathcal{F}(x,\qp^2,eB)$ can be written in a compact form as given in Eq.(\ref{sum_pp}). 
Finally, after regularization in the $\overline{MS}$ scheme,  the vacuum part of self-energy for the $pp$ loop with  $\vec{\,\,\,\,\qpr}=0$ becomes
\bea
(\Pi^{}_0)^{pp}_{\mbox{vac}}&=&g^2_{\pnn}\frac{m^2\qp^2 }{2\pi^2}\int_0^1 dx\Big[\psi\Big(\frac{\D}{2eB}\Big)+\frac{eB}{\D}\Big].
\label{pp_selfenergy}
\eea
Note that it contains  the pure vacuum part i.e one without explicit $eB$ dependence as well as the explicit magnetic field dependent vacuum contribution.
The vacuum contribution from $nn$ loop is given by 
\bea
(\Pi^{}_0)^{nn}_{\mbox{vac}}&=&-ig^2_{\pnn}\int\frac{d^4k}{(2\pi)^4}\mbox{Tr}\Big[\g5\slq S_n(k)\g5\slq S_n(p=q+k)\Big]\nn\\
&=&-i\int\frac{d^4k}{(2\pi)^4}\frac{N(k)}{(k^2-m^2+i\epsilon)(p^2-m^2+i\epsilon)}\nn\\
\mbox{where}\hspace{0.5cm}N(k)&=&4g^2_{\pnn}\Big[2(k\cdot q)^2+q^2(k\cdot q-k^2-m^2)\Big].
\eea
This is a divergent integral and the momentum integration can be performed after standard Feynman parametrization. After regularization in $\overline{MS}$ scheme
the finite part of  the vacuum self-energy for the $nn$ loop can be obtained as
\bea
(\Pi^{}_0)^{nn}_{\mbox{vac}}&=&g^2_{\pnn}\frac{m^2\qp^2}{2\pi^2}\int_0^1 dx\, \ln \Big(\frac{\D}{\mu}\Big)\hspace{0.5cm} \mbox{where}\,\, \vec{\,\,\,\,\qpr}=0
\label{nn_selfenergy}
\eea
and $\mu$ is the scale of the theory having dimension of square mass.  At zero magnetic field,
the self-energy contribution of the $pp$ loop must coincide  with the contribution from the $nn$ loop in isospin symmetric matter.  Now since 
the complete form of the propagators is used in order to derive  Eq.(\ref{pp_selfenergy}), we obtain a non-perturbative result 
as long as expansion in terms of  $eB$ is concerned. For that reason, we can not simply put $eB=0$ there to obtain the zero field contribution. Instead, 
we  obtain  a perturbative expansion of the $pp$ result around $eB=0$. The $eB\rightarrow 0$ expansion of Eq.(\ref{pp_selfenergy}) neglecting 
$\mathcal{O}((eB)^2)$ term is given  by 
\bea 
(\Pi^{}_0)^{pp}_{\mbox{vac}}&=&g^2_{\pnn}\frac{m^2\qp^2}{2\pi^2}\int_0^1 dx\, \ln \Big(\frac{\D}{2eB}\Big)\nn\\
&=&g^2_{\pnn}\frac{m^2\qp^2}{2\pi^2}\int_0^1 dx\, \Big[\ln \Big(\frac{\D}{\mu}\Big)-\ln \Big(\frac{2eB}{\mu}\Big)\Big].\nn\\
\label{ppeB0}
\eea
The first term in the square bracket exactly matches with  Eq.(\ref{nn_selfenergy}) whereas the second term diverges at $eB=0$. Hence, to match the 
two expressions identically i.e  irrespective of the value of external momentum 
we must modify Eq.(\ref{pp_selfenergy}) as,
\bea 
(\Pi^{}_0)^{pp}_{\mbox{vac}}&=&g^2_{\pnn}\frac{m^2\qp^2 }{2\pi^2}\int_0^1 dx\Big[\psi\Big(\frac{\D}{2eB}\Big)+\frac{eB}{\D}+\ln \Big(\frac{2eB}{\mu}\Big)\Big].
\eea 
By  demanding identical contributions from $pp$ and $nn$ loop at vanishing magnetic field, one in fact imposes here a physical condition to extract out the 
finite part of the self energy.    

We now turn to the thermal contribution.
The real part of the thermal contribution for proton-proton($pp$) loop can  be obtained following a similar  procedure employed  in case of charged pions and  is  given by 
\bea
\mathcal{R}\mbox{e}(\Pi^{pp}_0)_\eta&=&\sum_{n,l=0}^{\infty}\int\frac{dk_z}{(2\pi)^3}\mathcal{P}\Big[\frac{\mathcal{N}^{n,l}_{||}(k^0=-q^0
+\omega^l_p)n^{+}_{p,l}}{2\omega^l_p\{(q^0-\omega^l_p)^2-(\omega^n_k)^2\}}
+\frac{\mathcal{N}^{n,l}_{||}(k^0=-q^0-\omega^l_p)n^{-}_{p,l}}{2\omega^l_p\{(q^0+\omega^l_p)^2-(\omega^n_k)^2\}}\nn\\
&+&\frac{\mathcal{N}^{n,l}_{||}(k^0=\omega^n_k)n^{+}_{k,n}}{2\omega^n_k\{(q^0+\omega^n_k)^2-(\omega^l_p)^2\}}+
\frac{\mathcal{N}^{n,l}_{||}(k^0=-\omega^n_k)n^{-}_{k,n}}{2\omega^n_k\{(q^0-\omega^n_k)^2-(\omega^l_p)^2\}}\Big]\,\,\,\,\,\,{\rm where}\nn\\
\mathcal{N}^{n,l}_{||}(k)&=&4\pi g^2_{\pnn}eB\Big[4eB\qp^2n\delta_{n-1,l-1}+\{2(\qp\cdot\kp)^2+\qp^2(-\kp^2-m^2+\qp\cdot\kp)\}(\delta_{n,l}+\delta_{n-1,l-1})\Big]
\label{double_sum}
\eea
with $\omega_k^n=\sqrt{k_z^2+m^2+2neB}$ and $\omega_p^l=\sqrt{p_z^2+m^2+2leB}=\sqrt{(q_z+k_z)^2+m^2+2leB}$. Here $\mathcal{P}$ represents the principle value of the argument.
%\omega_k^n&=&\sqrt{k_z^2+m^2+2neB}\nn\\
%\omega_p^l&=&\sqrt{p_z^2+m^2+2leB}\nn\\
%&=&\sqrt{(q_z+k_z)^2+m^2+2leB}\nn\\
%\eea
 In case of  neutron-neutron($nn$) loop
\bea
\mathcal{R}\mbox{e}(\Pi^{nn}_0)_\eta&=&\int\frac{d^3k}{(2\pi)^3}\mathcal{P}\Big[\frac{\mathcal{N}(k^0=-q^0+\omega_p)n^{+}_{p}}{2\omega_p\{(q^0-\omega_p)^2-\omega_k^2\}}+\frac{\mathcal{N}(k^0=-q^0-\omega_p)n^{-}_{p}}{2\omega_p\{(q^0+\omega_p)^2-\omega_k^2\}}\nn\\
&+&\frac{\mathcal{N}(k^0=\omega_k)n^{+}_{k}}{2\omega_k\{(q^0+\omega_k)^2-\omega_p^2\}}+
\frac{\mathcal{N}(k^0=-\omega_k)n^{-}_{k}}{2\omega_k\{(q^0-\omega_k)^2-\omega_p^2\}}\Big]\hspace{0.5cm}{\rm where}\nn \\
\mathcal{N}(k)&=&4 g^2_{\pnn}\Big[2(\qp\cdot\kp)^2+\qp^2(\qp\cdot\kp-k^2-m^2)\Big]
\eea
with $\omega_k=\sqrt{\vec{k}^{2}+m^2}$ and $\omega_p=\sqrt{(\vec{q}+\vec{k})^2+m^2}$.  It should be mentioned here that in case of $pp$ loop, the expression is
obtained with the simplifying assumption that  $\vec{\,\,\,\,\qpr}=0$ so that the $\kpr$ integral can be performed exactly using the orthogonality condition of 
generalized Laguerre polynomials which renders the products of Laguerre polynomials into simple Kronecker Deltas and  in turn makes
it trivial to convert the double summation structure of the self-energy  into a single sum over Landau Levels. Obviously,  the same assumption does not
provide any such simplification for $nn$ loop. In this case also  $\vec{\,\,\,\,\qpr}=0$ is taken for consistency. 
Thus, the one loop  vacuum self-energy of neutral pion becomes
\bea
(\Pi^{}_0)_{\mbox{vac}}&=&(\Pi^{}_0)^{pp}_{\mbox{vac}}+(\Pi^{}_0)^{nn}_{\mbox{vac}}\nn\\
&=&g^2_{\pnn}\frac{m^2\qp^2 }{2\pi^2}\int_0^1 dx\Big[\psi\Big(\frac{\D}{2eB}\Big)+\frac{eB}{\D}+\ln \Big(\frac{2eB}{\mu}\Big)+\ln \Big(\frac{\D}{\mu}\Big)\Big]\nn\\
&=&g^2_{\pnn}\frac{m^2\qp^2 }{2\pi^2}\int_0^1 dx\Big[\psi\Big(\frac{\D}{2eB}\Big)+\frac{eB}{\D}+\ln \Big(\frac{2eB}{\D}\Big)\Big]+2g^2_{\pnn}
\frac{m^2\qp^2 }{2\pi^2}\int_0^1 dx
\Big[\ln \Big(\frac{\D}{\mu}\Big)\Big]\nn\\
&=&(\Pi^{}_0)_{\mbox{vac}}(eB\ne 0)+(\Pi^{}_0)_{\mbox{vac}}(eB=0).
\eea

Unlike the charged pion case, the external transverse momentum of the neutral pions is continuous.
To obtain the effective mass  of the neutral pions as a function of $T$, $\mu_N$ and $eB$, we  solve the DSE   given by 
\bea
m^{{\ast}2}_{\pi^0}-m_{\pi^0}^2+\mathcal{R}\mbox{e}\Pi_{0}(m^{{\ast}2}_{\pi^0},\vec{q}=0,eB)=0
\label{dse}
\eea
where $\mathcal{R}\mbox{e}\Pi_{0}$ contains the sum of explicit $eB$ dependent vacuum and thermal contribution and  
 the renormalized pion mass in vacuum $m_{\pi^0}$ is  taken to be same as $m_{\pi^{\pm}}^{}$ and will be denoted as $m_{\pi}$ in subsequent sections.

%%%%%%%%%%%%%%%%%%%%%%%%%%%%%%%%%%%%%%%%%%%%%%%%%%%%%%%%%%%%%%%%%%%%%%%%%%%%%%%%%%%%%%%%%%%%%%%%%%%%%%%%%%%%%%%%%%%%%%%%%%%%%%%%%%%%%%%%%%%%%%%%%%%%%%%%%%%%%%%%%%%%

\section{Results and discussions}
\begin{center}
\begin{figure}
 \subfloat[]{\includegraphics[width=2.5in,angle=-90]{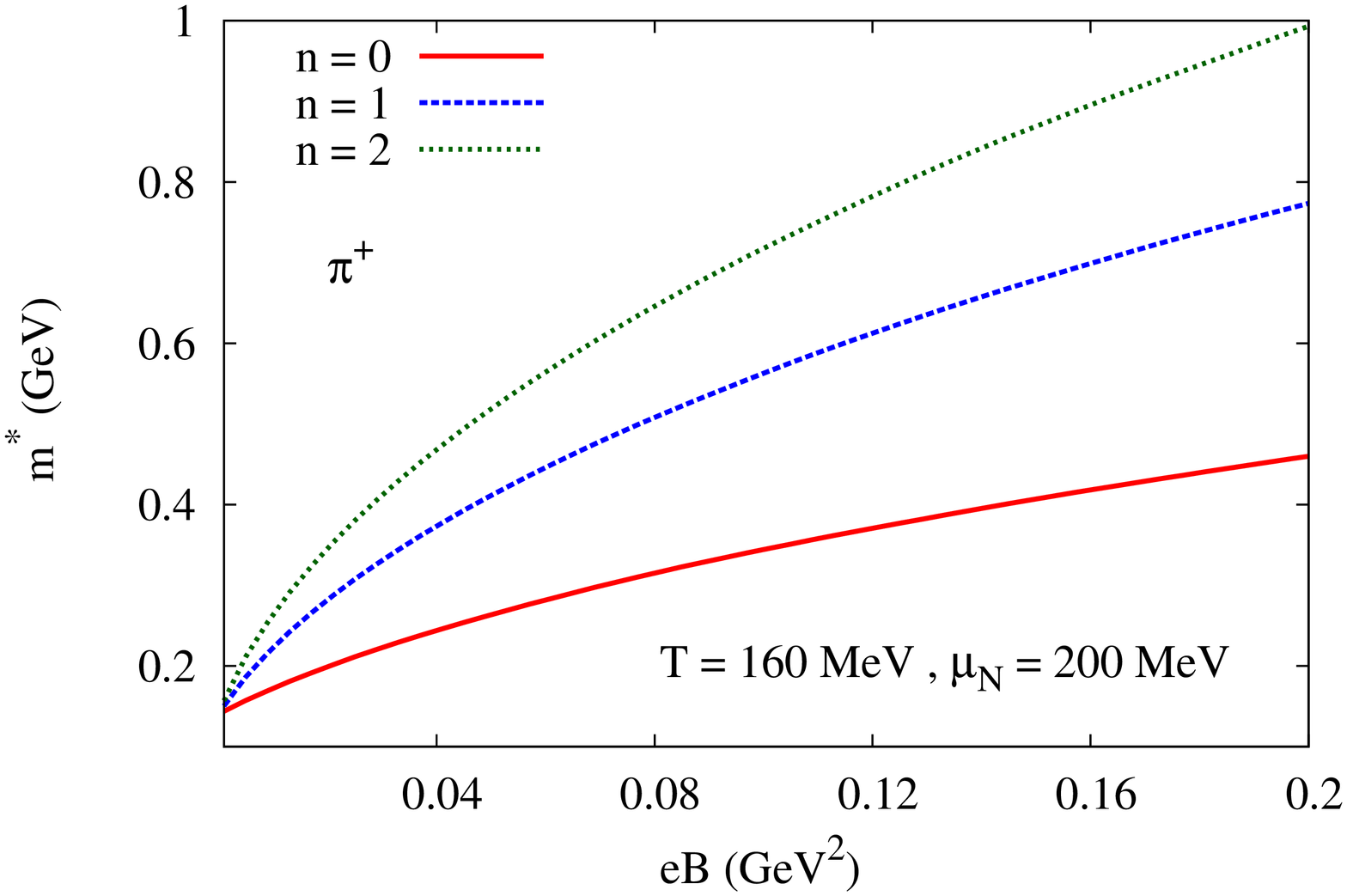}}
 \subfloat[]{\includegraphics[width=2.5in,angle=-90]{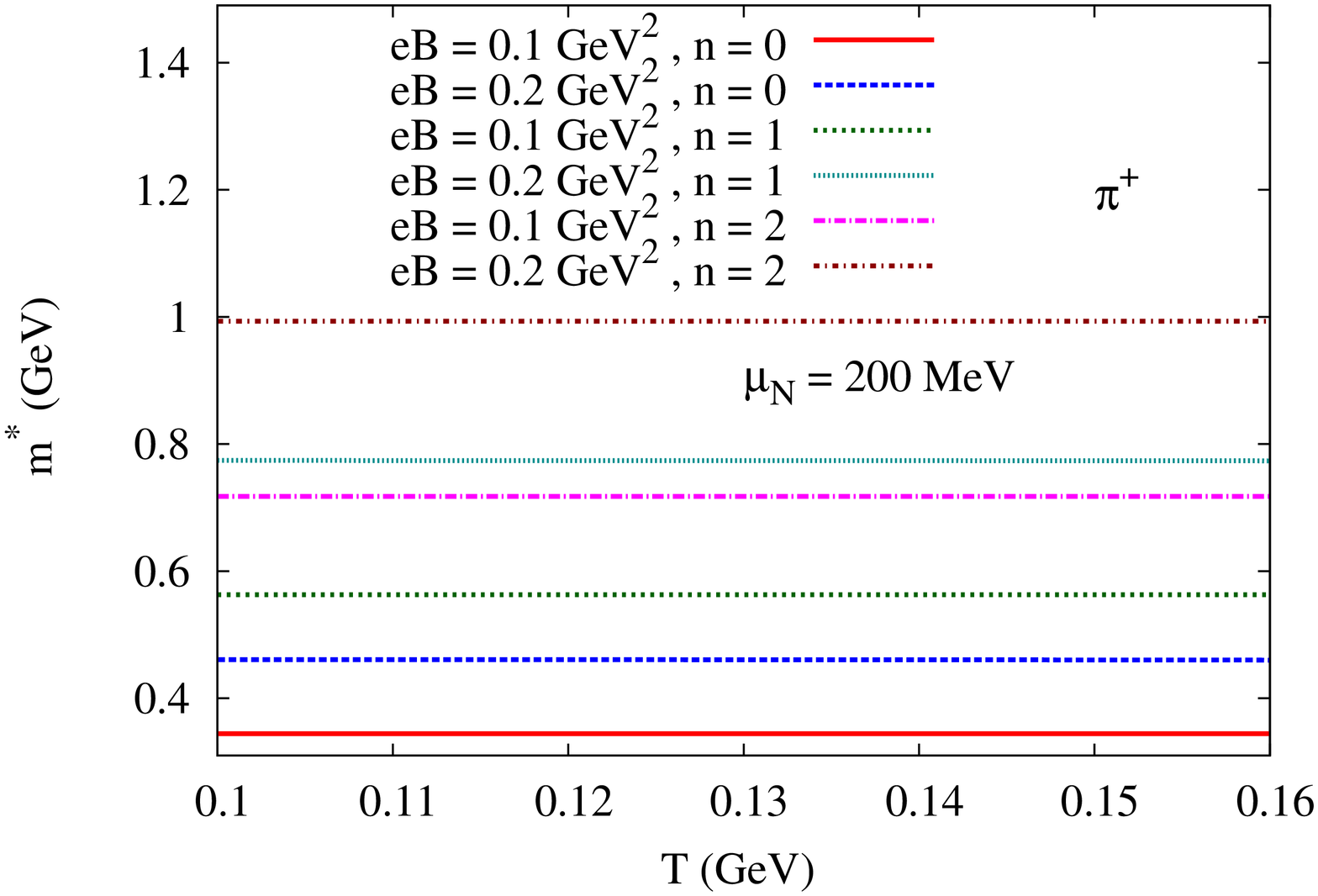}}\\
 \subfloat[]{\includegraphics[width=2.5in,angle=-90]{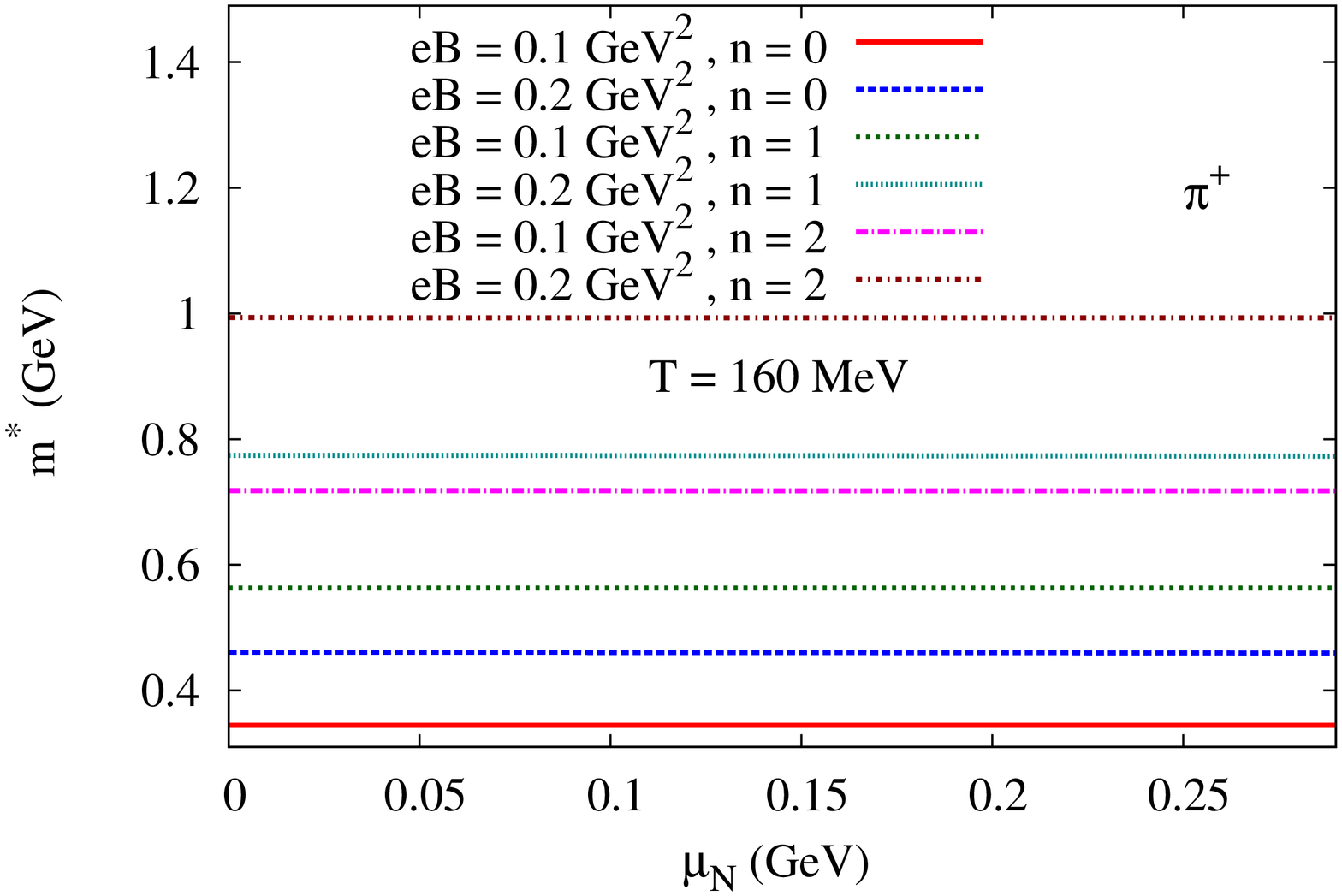}}  
 \caption{ $eB$ dependence of the effective mass  due to the  
 magnetic field dependent  vacuum contribution  and the  thermal 
 contribution  of the self-energy along with the  Landau shift  are shown at  T=160 MeV and $\mu_{N}=200$ MeV. 
 Effective mass variations of  $\pi^+$ with respect to $T$ and $\mu_N$ are  plotted for $eB=0.1$ and 0.2 GeV$^2$.}
\label{mstarpm}
\end{figure}
\end{center}

\begin{figure}[]
 \subfloat[]{\includegraphics[width=2.5in,angle=-90]{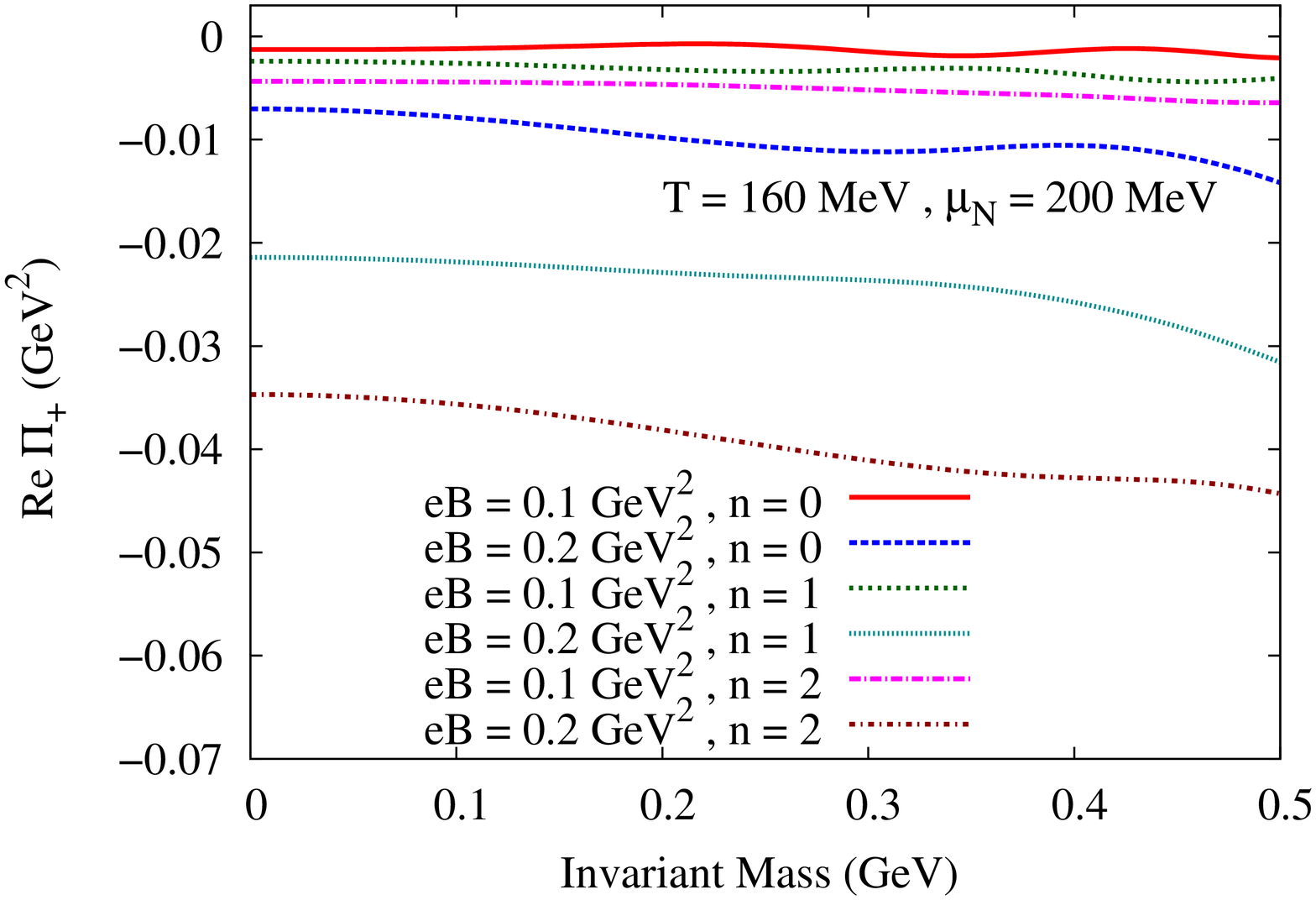}}
 \subfloat[]{\includegraphics[width=2.5in,angle=-90]{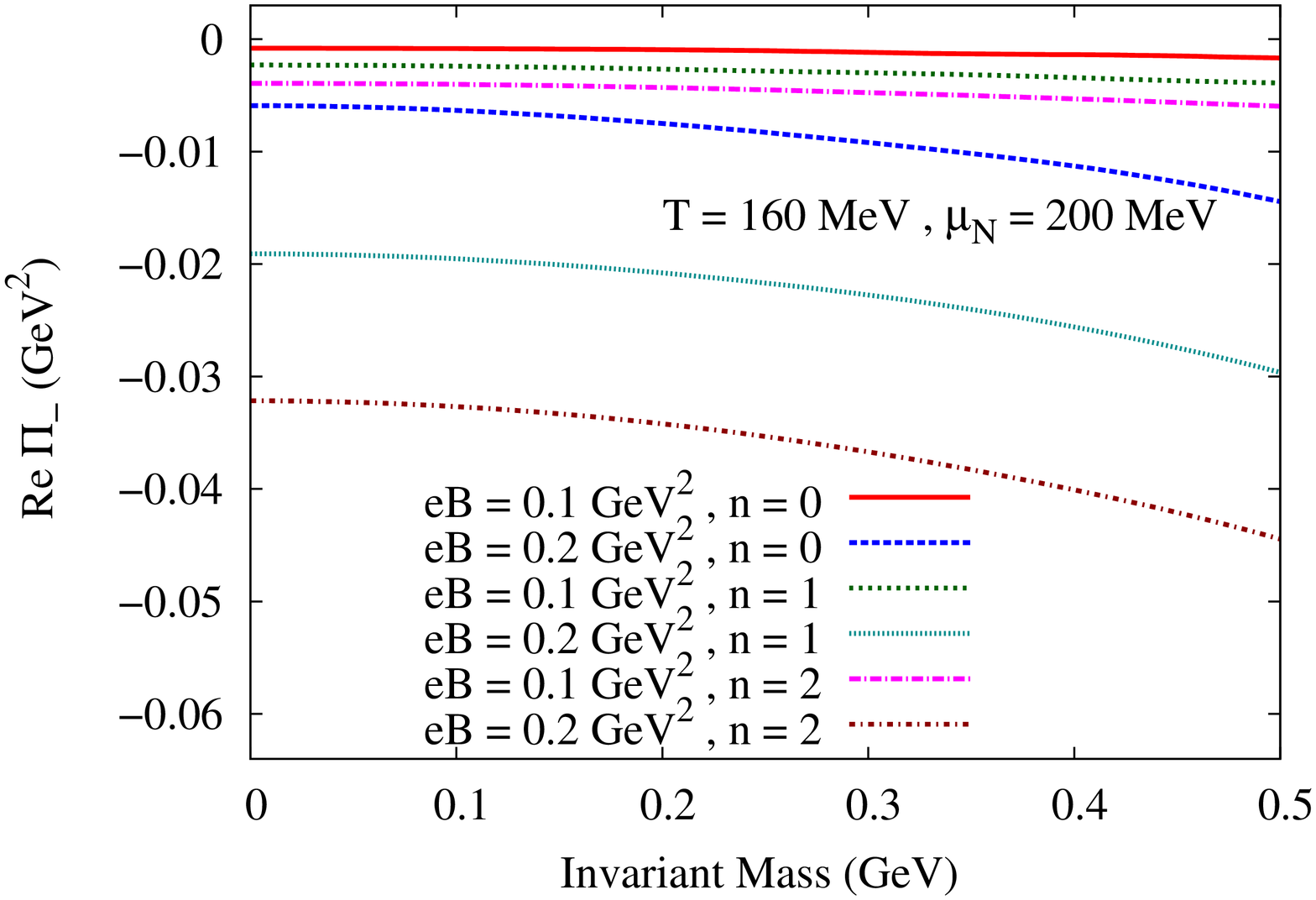}}
\caption{Real parts of the thermal contributions are plotted as a function of $q^0$ for two different values of chemical potential, $\mu_N$= 0 and 200 MeV with 
 $|q_z|$=200 MeV. External parameters $eB$ and T are set at 0.1 GeV$^2$ and 160 MeV respectively.}
 \label{reprtpm}
\end{figure}
  
In this section we present the numerical results obtained by solving the DSE given in Eq.(\ref{dse}). We have taken $f^2_{\pnn}/{4\pi}=0.0778$  and $m_\pi$=0.14 GeV.
The nucleon mass is taken as 0.938 GeV. In case of evaluating the effective masses, we have summed up to 300 Landau levels of the loop particles. For stronger magnetic fields i.e 
$eB>0.1$ GeV$^2$,the  
results are found to converge at much lower value of the maximum Landau Level used. However, for $eB\rightarrow 0$, more than  200 landau levels 
are needed to produce a convergent numerical result. In case of charged pions we have taken the scale $\mu$ as the square of the neucleon mass.
At first we solve the DSE for the charged pions given by 
\bea
m_{}^{*\,2}-m_{\pi}^{\,2}+\mathcal{R}\mbox{e}\Pi_{\pm}(m^{*\,2},\qpr^2=(2n+1)eB,eB)-(2n+1)eB=0
\label{dsepm1}
\eea
where $m_{}^{*}=m_{\pi^+}^{*}(m_{\pi^-}^{*})$ when $\mathcal{R}\mbox{e}\Pi_{+}(\mathcal{R}\mbox{e}\Pi_{-})$ is used. The variation of $m_{}^{*}$ with $eB$, $T$ and 
$\mu_N$ for  $\pi^+$ is shown in Fig.\ref{mstarpm}. It can be seen from  Fig.\ref{mstarpm}(a) that the plots are completely dominated by the trivial Landau quantization 
of the external pion. In other words, for each value of Landau Level, the linear $(2n+1)eB$ term present in Eq.(\ref{dsepm}) affects 
the effective mass much more  compared to the one loop self-energy correction. Moreover, as the temperature and $\mu_N$ dependence of the
effective mass comes only through the self-energy, one can  expect that in the $eB$ dominated scale, they will be insignificantly small. Accordingly 
$m^\ast$ seems to be independent of the variation of $T$ and $\mu_N$ as shown in (b) and (c) part of Fig.\ref{mstarpm}. It should be mentioned here that 
as the DSE of  $\pi^-$ is different from that of $\pi^+$ only 
in the expression of $\mathcal{R}\mbox{e}\Pi$, the $m^\ast$ plots for $\pi^-$ will be exactly superimposed on those of $\pi^+$. However, differences
between the two charged species  can be  observed  in Fig.(\ref{reprtpm}) where  the variation of the real part of the self-energy with the invariant mass is shown
for two different values of $eB$ with first three Landau Levels. Although both of the self-energies decrease with invariant mass and remain negative throughout, the
small oscillatory behaviour present in case of $\mathcal{R}\mbox{e}\Pi_{+}$ can not be observed for $\mathcal{R}\mbox{e}\Pi_{-}$. To unveil the contribution 
of the real part of the self-energy correction, now we neglect the trivial shift and solve 
\bea
m^{{\ast}2}_{\pi^{\pm}}-m_{\pi}^2+\mathcal{R}\mbox{e}\Pi_{\pm}(m^{{\ast}2}_{\pi^\pm},\vec{q}=0,eB)=0.
\label{dse_subtracted}
\eea
In other words, the  effective mass is measured with respect to the trivial Landau shift. This kind of situation may occur in principle  when different species
of charged particles are present in the system and all of which undergo equivalent trivial Landau shifts. In that case  the real part of the self-energies 
will play the deciding role in the characterization of the effective masses. Moreover, comparing Eq.(\ref{dse}) and  Eq.(\ref{dse_subtracted}), one can 
observe that Eq.(\ref{dse_subtracted}) brings down the charged pions in equal footing with the neutral pions only in the difference of the self-energy. 
    
\begin{figure}
\begin{minipage}{\textwidth}
 \subfloat[]{\includegraphics[width=2.5in,angle=-90]{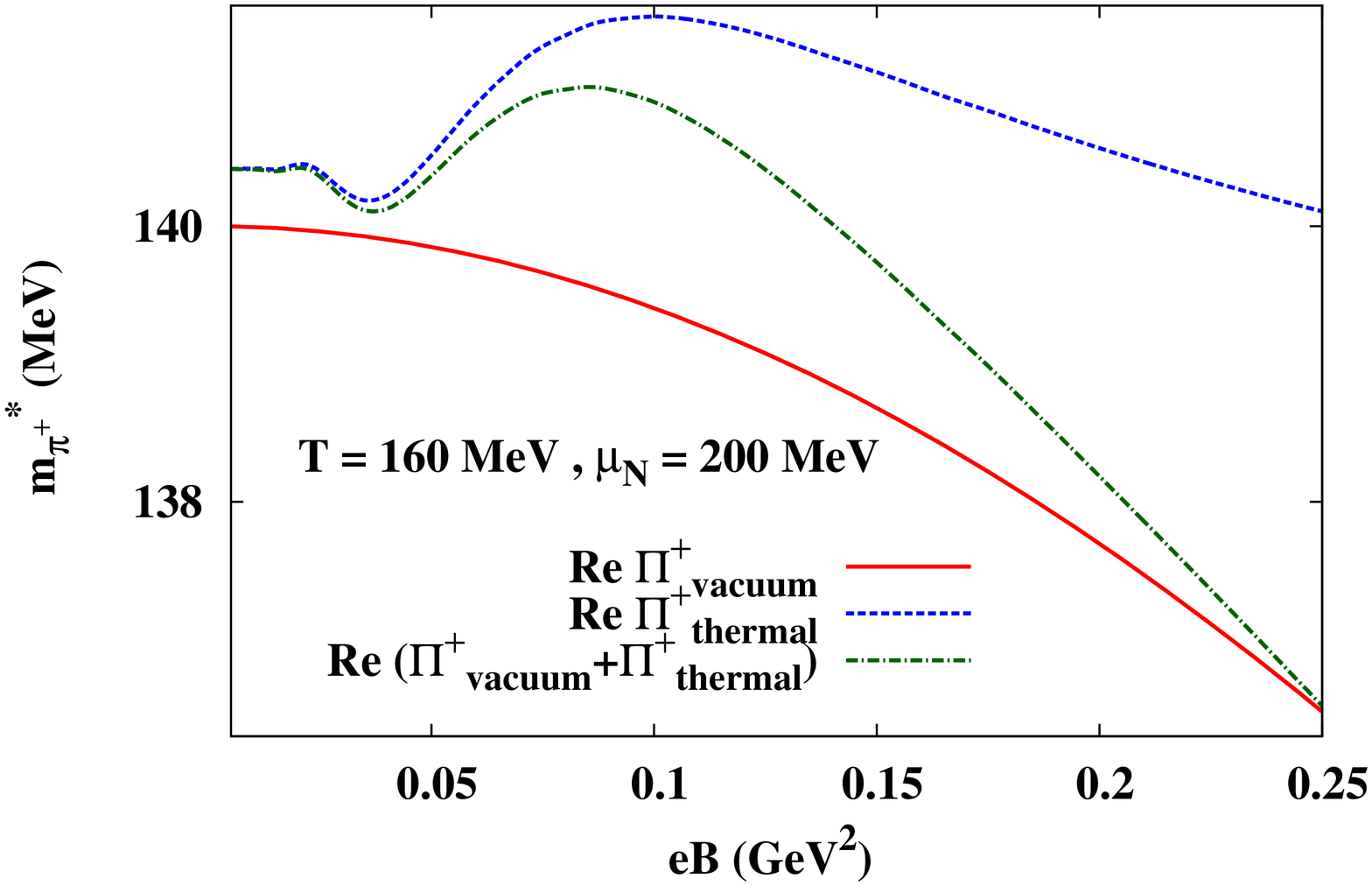}}
 \subfloat[]{\includegraphics[width=2.5in,angle=-90]{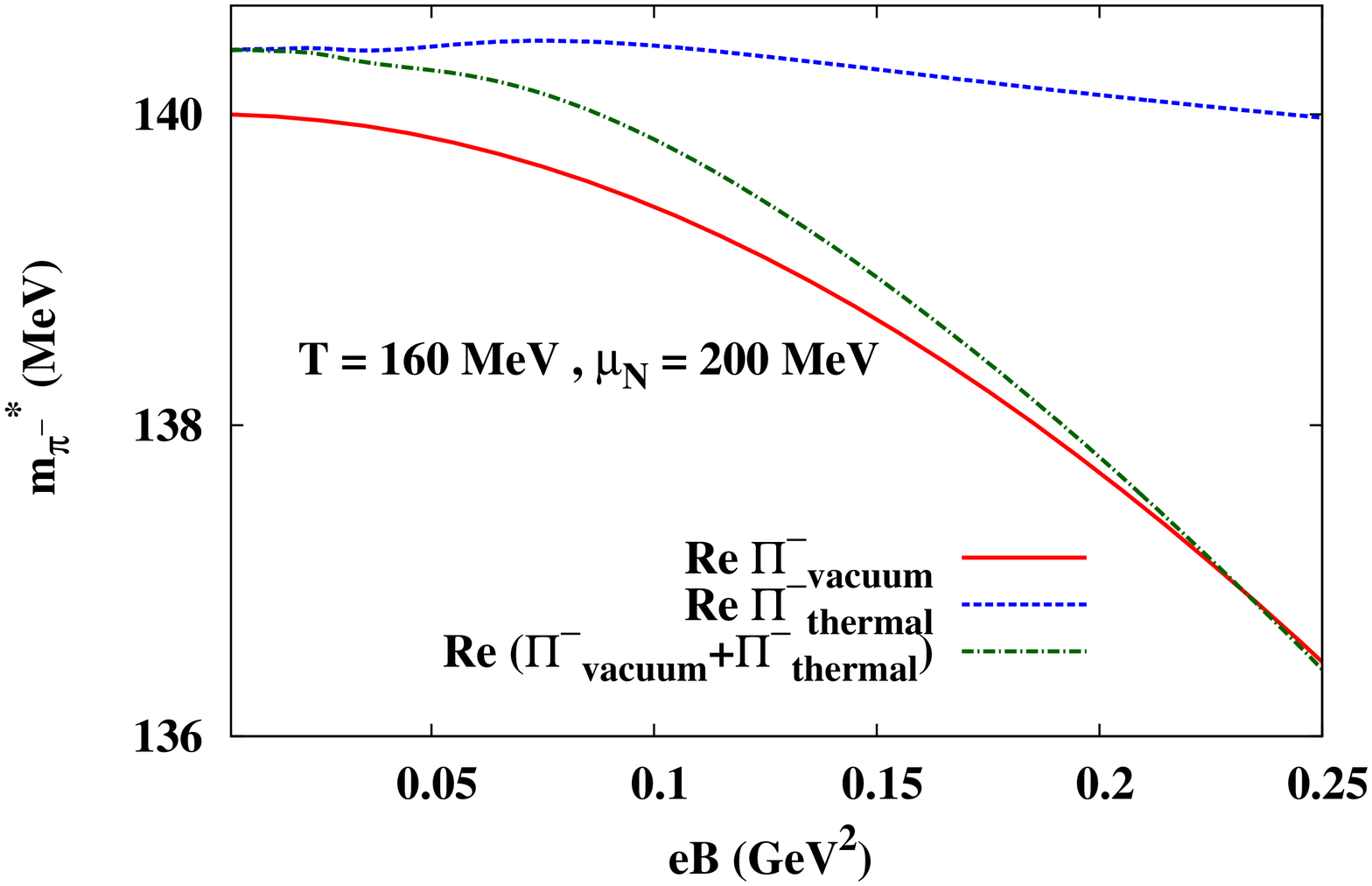}}\\
  \subfloat[]{\includegraphics[width=2.5in,angle=-90]{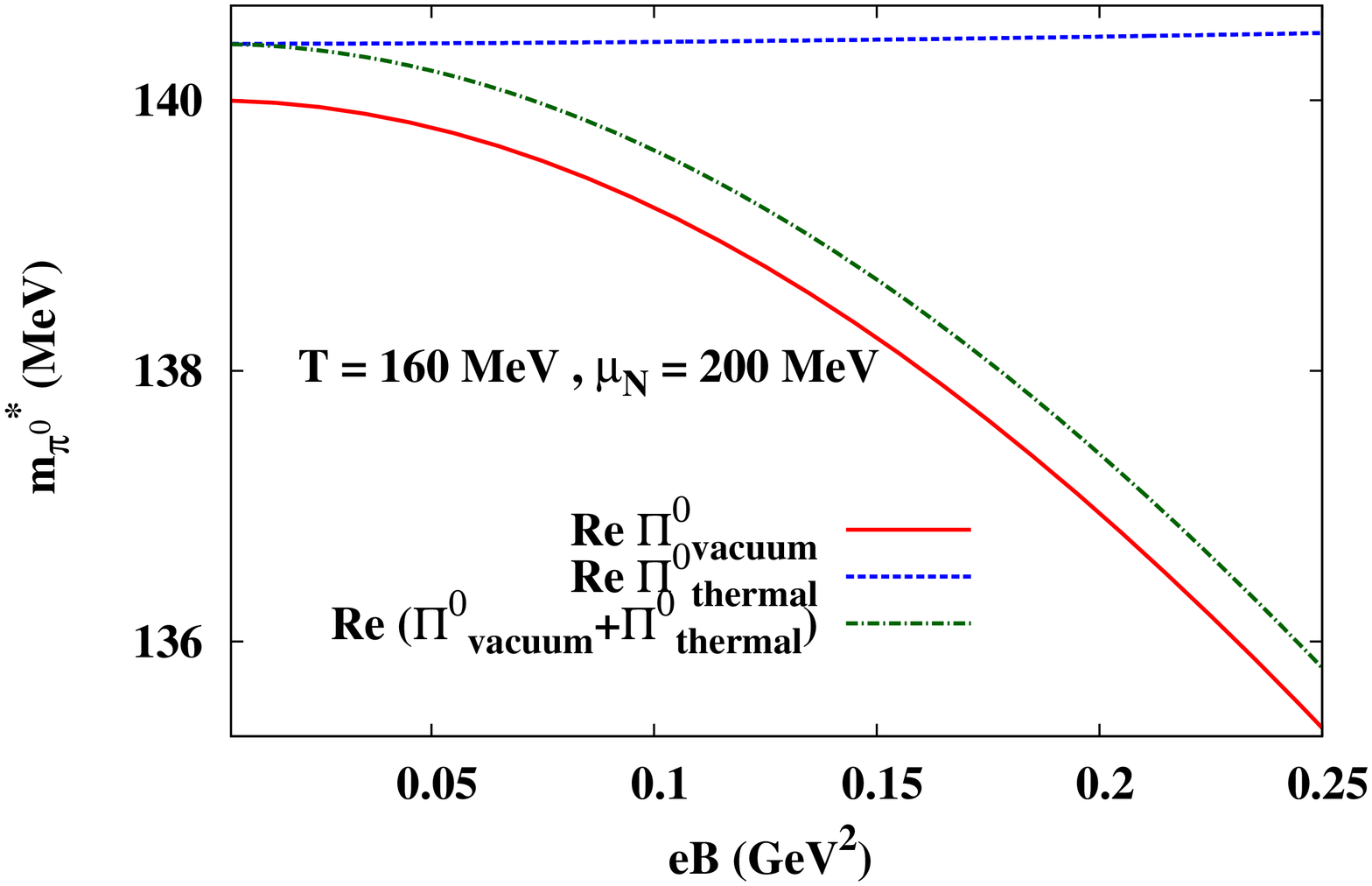}}
 \caption{ $eB$ dependence of the effective mass  due to the  
 magnetic field dependent  vacuum contribution  and the  thermal 
 contribution  of the self-energy  are shown at  T=160 MeV and $\mu_{N}=200$ MeV. 
 Effective mass variations for  the charged pions are plotted in the upper panel whereas 
 the lower panel describes the neutral pions. The total self-energy contribution is also shown in each case for comparison.}
\label{drv}
\end{minipage}
%\end{figure}
%\end{center}
%\begin{figure}[htb]
\begin{minipage}{\textwidth}
 \subfloat[]{\includegraphics[width=2.5in,angle=-90]{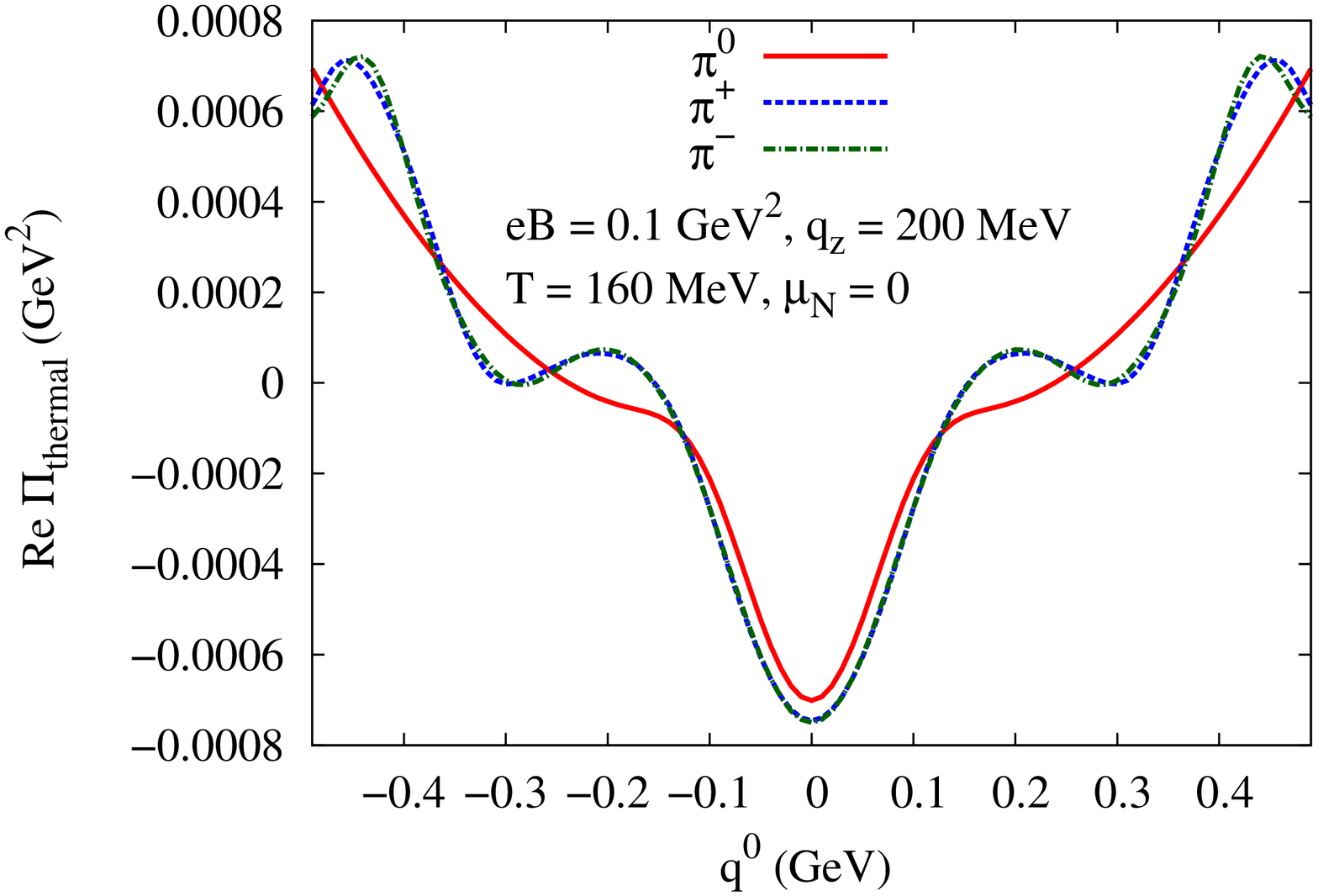}}
 \subfloat[]{\includegraphics[width=2.5in,angle=-90]{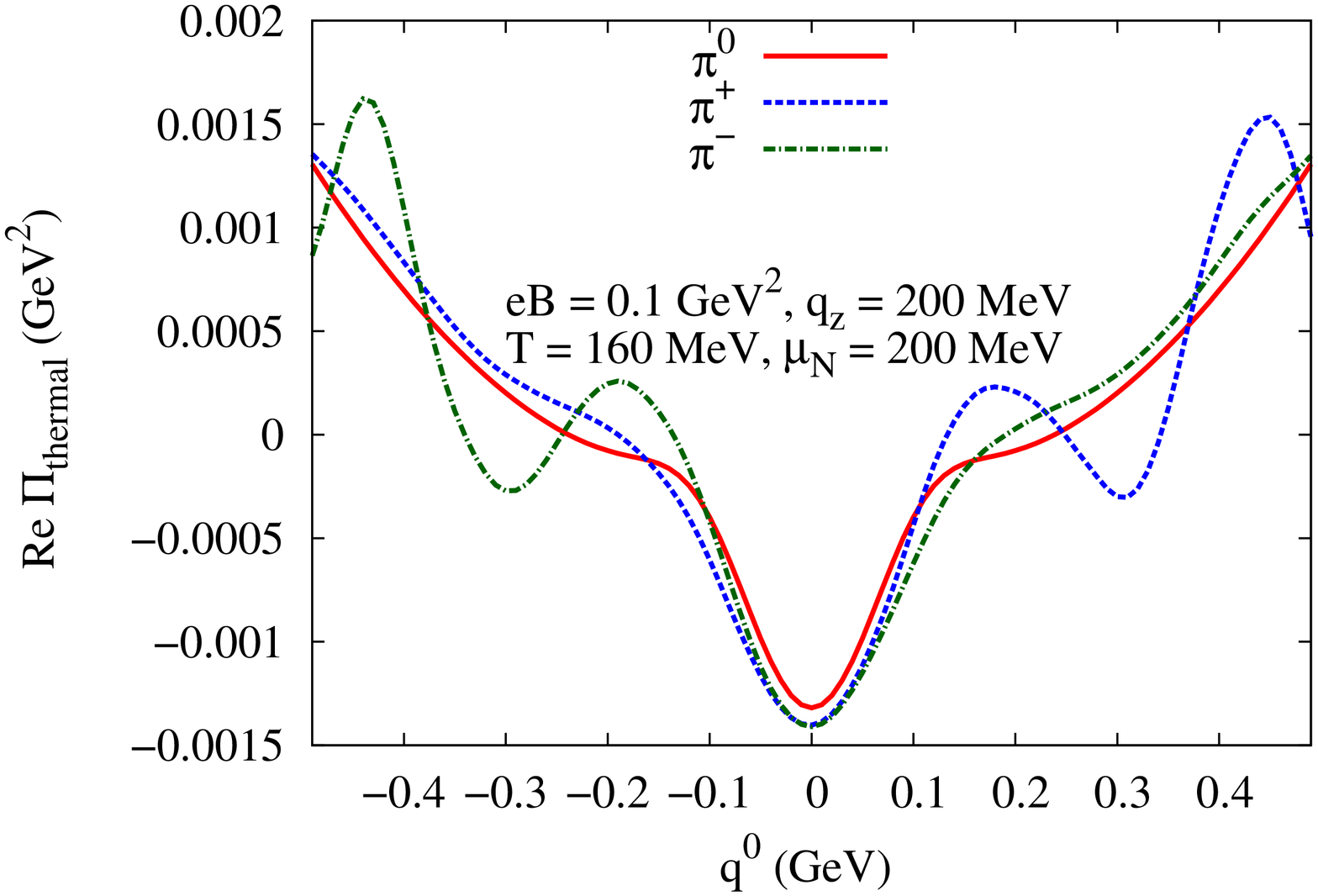}}
\caption{Real parts of the thermal contributions are plotted as a function of $q^0$ for two different values of chemical potential, $\mu_N$= 0 and 200 MeV with 
 $|q_z|$=200 MeV. External parameters $eB$ and T are set at 0.1 GeV$^2$ and 160 MeV respectively.}
 \label{reprt}
 \end{minipage}
\end{figure}

In Fig.\ref{drv}, the effective mass variation as a function of external magnetic field has been shown at a given temperature of 160 MeV and chemical 
 potential of 200 MeV. Considering  magnetic field dependent vacuum contribution and the  thermal contribution  of the self-energy separately, one can
 compare with the total contribution as shown. When  only  the thermal contribution is taken into account, it can be noticed that
 with the increase of external magnetic field, the effective mass of the charged
 pions develops smooth oscillations. However, no such oscillations have been observed for neutral pion which instead, shows marginal increase in the thermal  effective mass
 with $eB$. On the other hand,        
 the effective mass due to the magnetic field dependent vacuum part decreases monotonically with $eB$ for $\pi_0$ as well as for $\pi_{\pm}$. 
 However, the decreasing nature is more pronounced in case of neutral pion.  It is clear from the figure that the field dependent vacuum part 
 of the self-energy can influence the $eB$ dependence of the effective mass  significantly. One can  notice that, for neutral pions, even the qualitative
 nature of the $eB$ dependence of the effective mass changes i.e from a slowly increasing nature it becomes a decreasing function of $eB$ 
 due to the incorporation of the vacuum part. 
 
\begin{figure}[H]
\captionsetup[subfigure]{oneside,margin={0in,0in}}
\subfloat[]{\includegraphics[width = 2.5in,angle=-90]{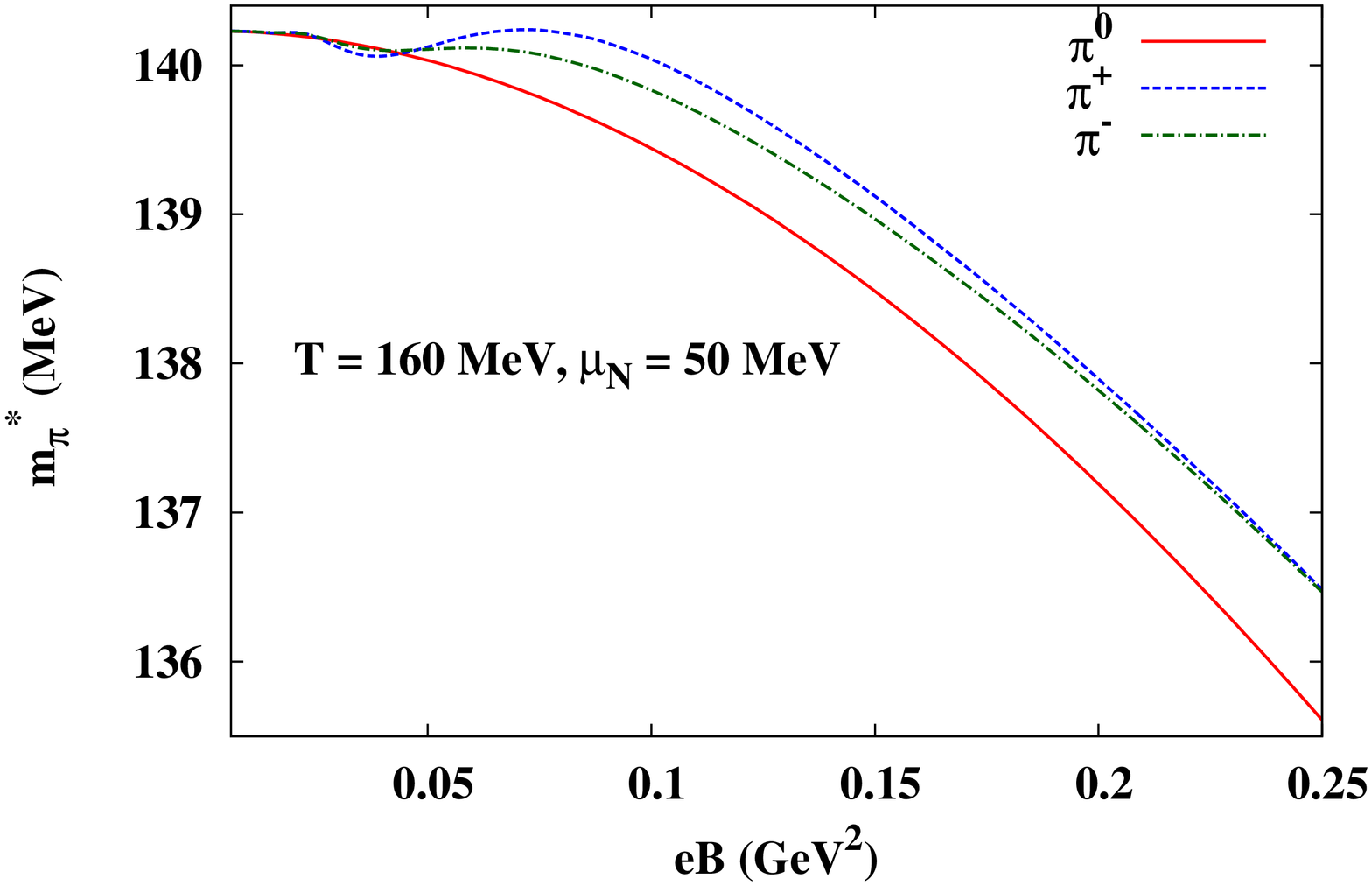}}\label{a} 
\subfloat[]{\includegraphics[width = 2.5in,angle=-90]{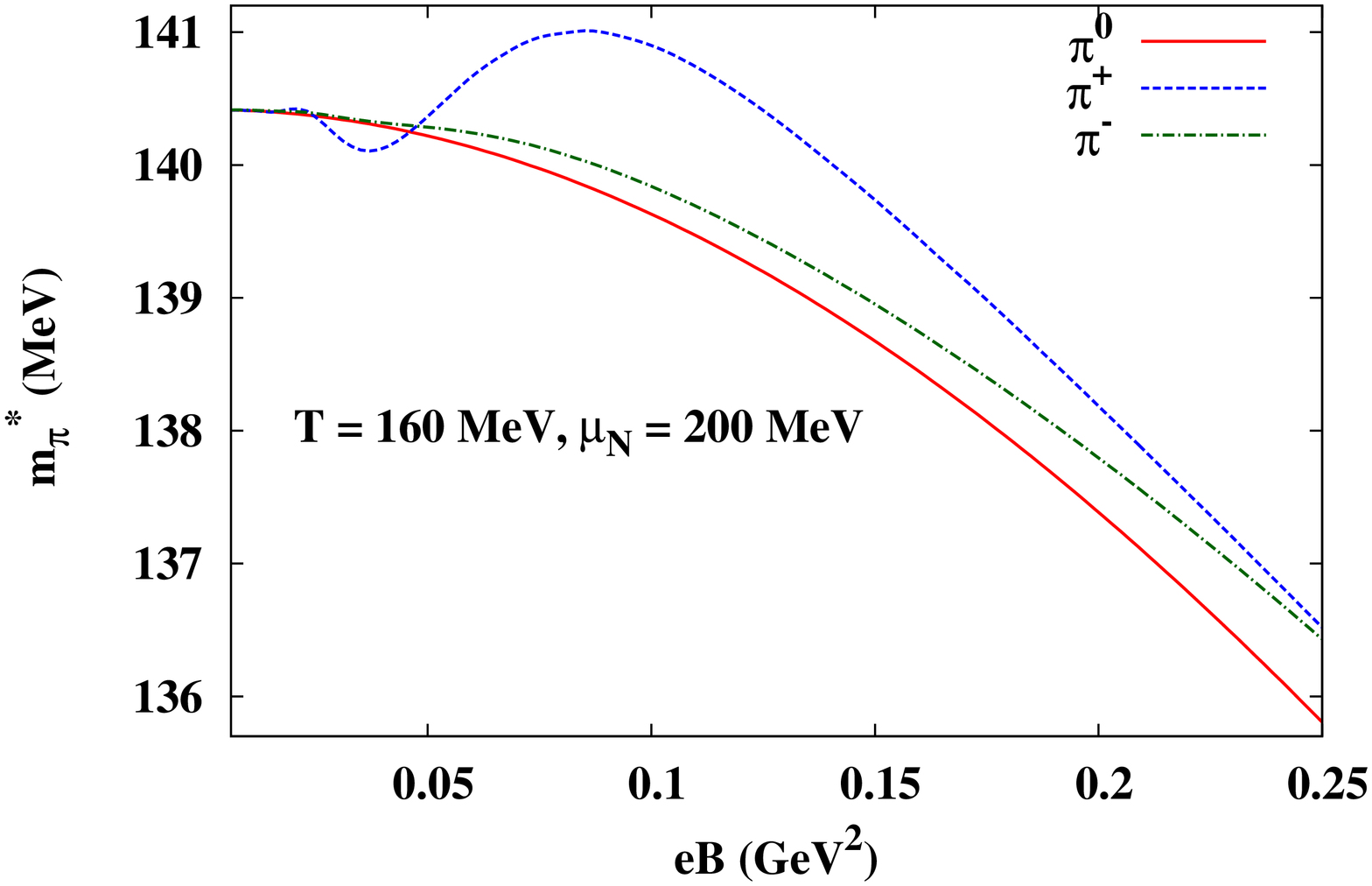}}\label{b}\\
\subfloat[]{\includegraphics[width = 2.5in,angle=-90]{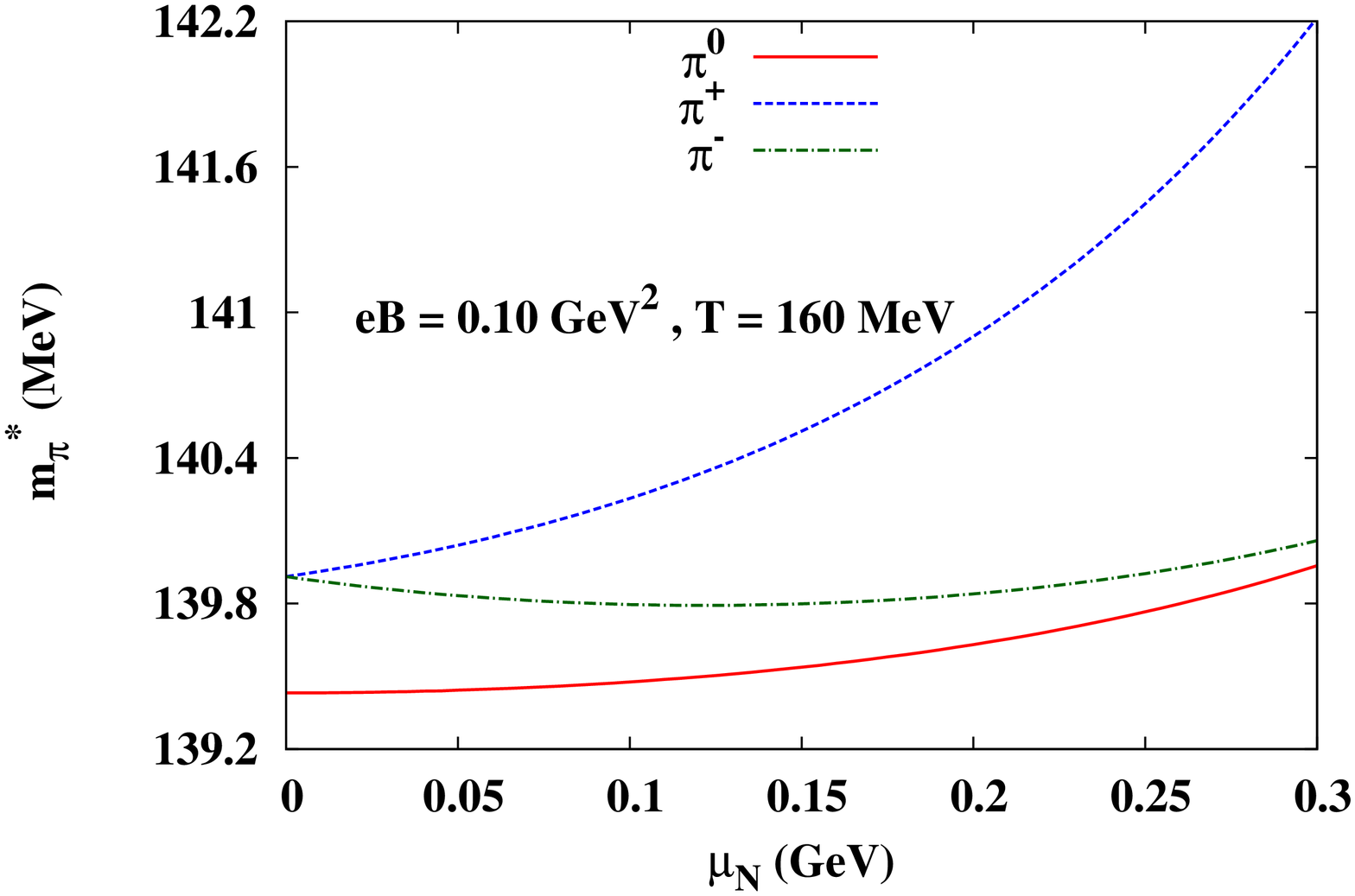}}\label{c}
\subfloat[]{\includegraphics[width = 2.5in,angle=-90]{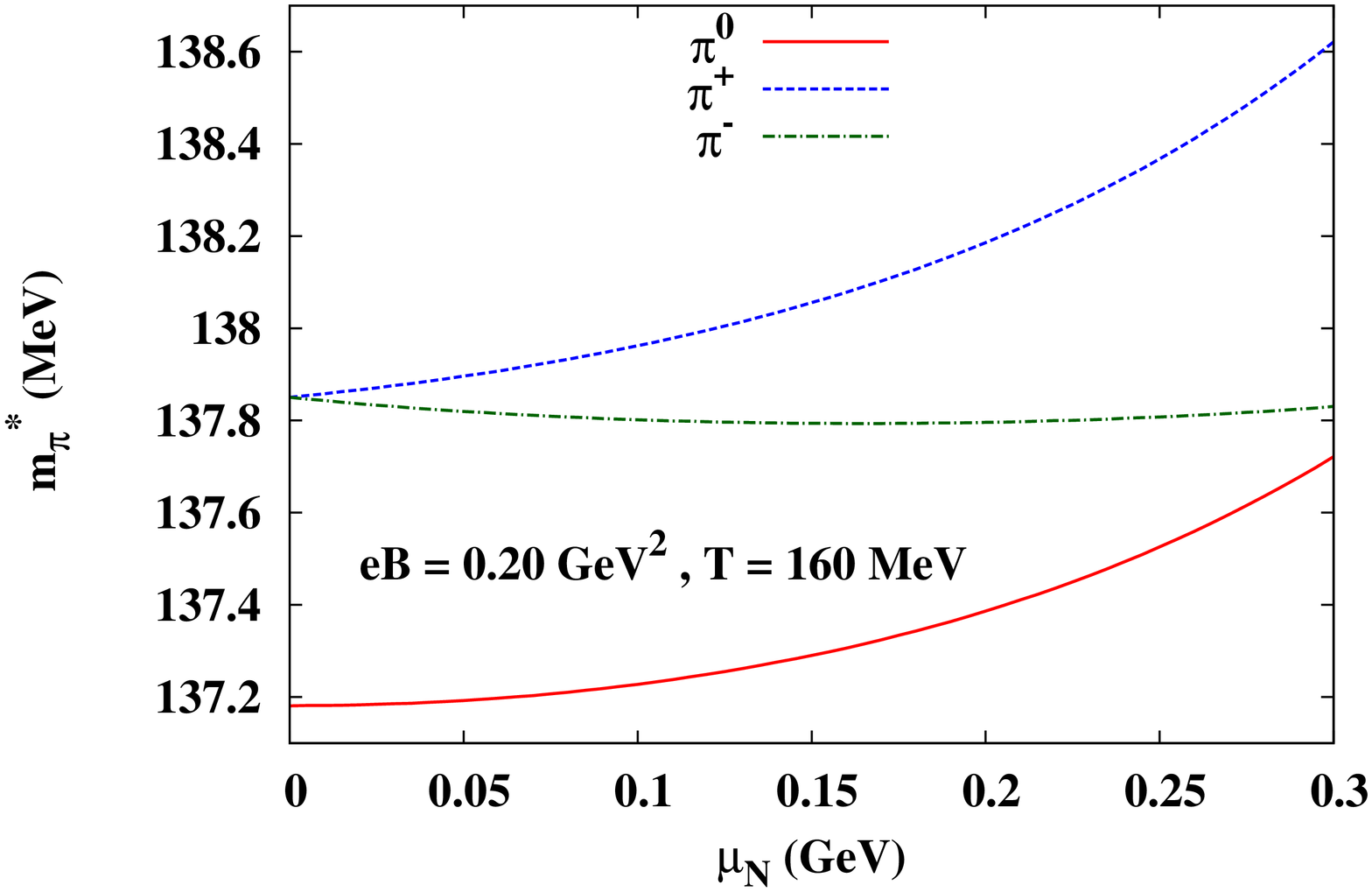}}\label{d}\\
\subfloat[]{\includegraphics[width = 2.5in,angle=-90]{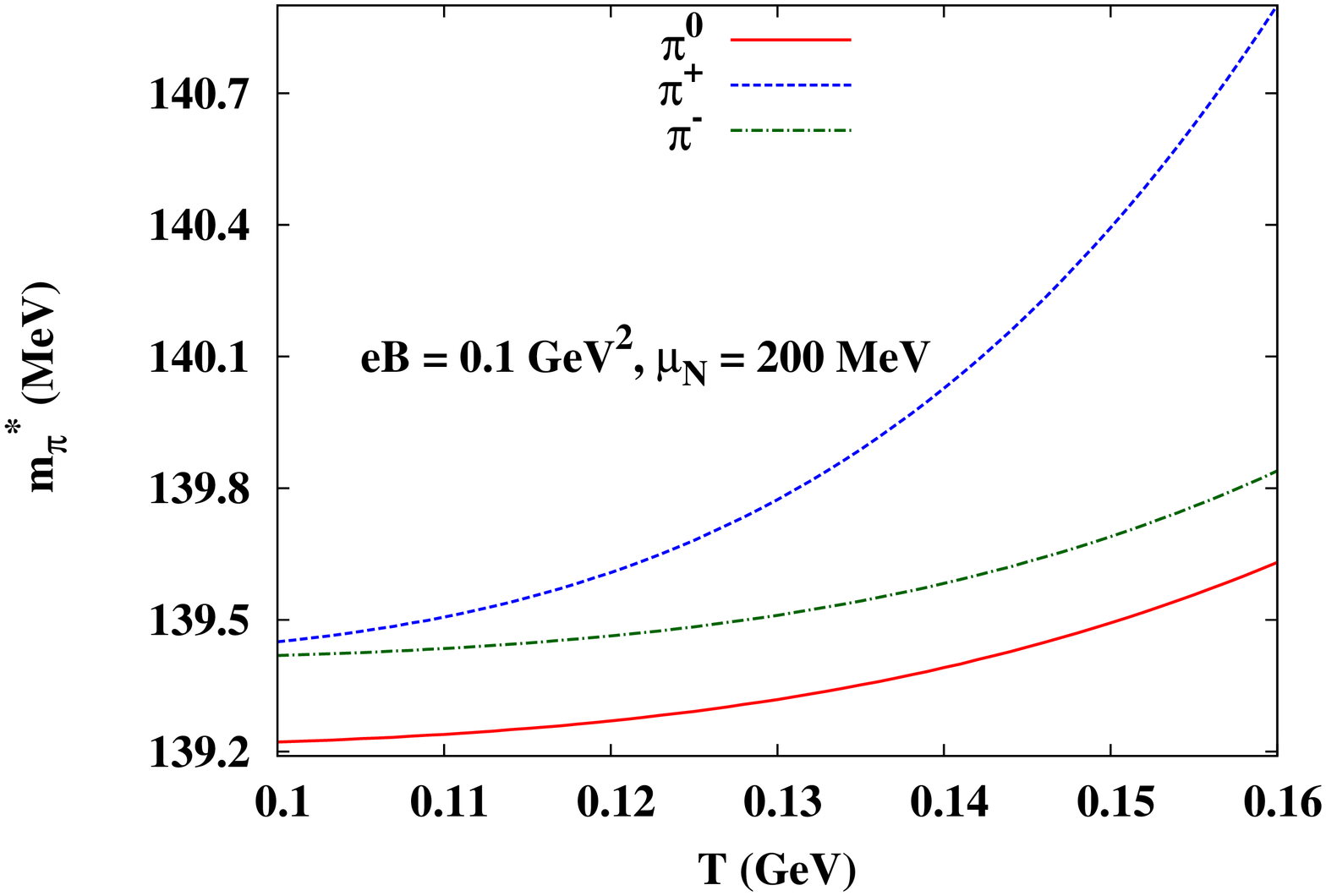}}\label{e}
\subfloat[]{\includegraphics[width = 2.5in,angle=-90]{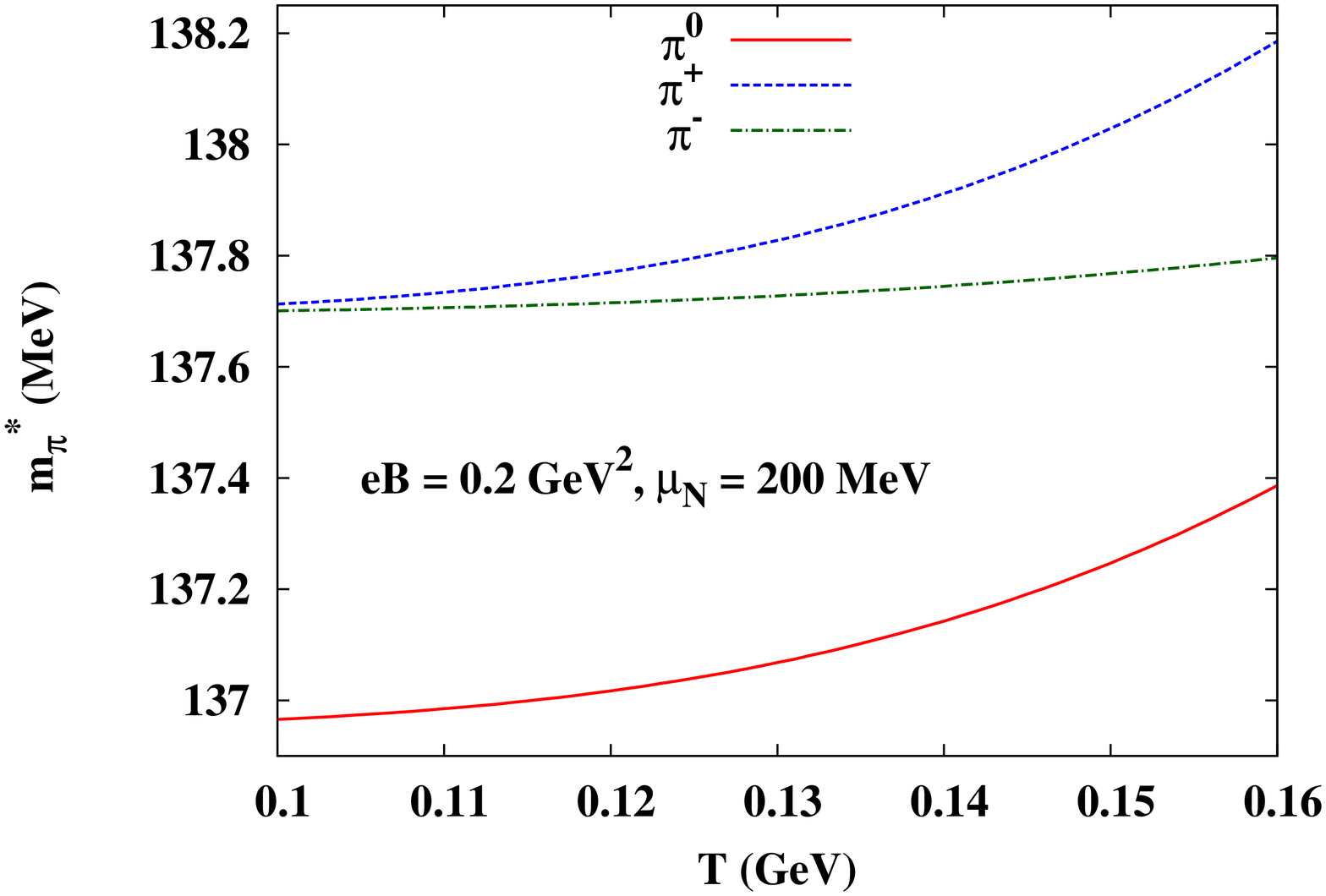}} \label{f}\\
\caption{Effective mass  variations with $eB$, T and $\mu_N$ are shown in the three horizontal panels respectively. First row contains 
two different plots corresponding to two different values of $\mu_N$(50 and 200 MeV respectively )with T=160 MeV. At the same temperature, 
second row describes the $\mu_N$ dependence for $eB$=0.10 and 0.20 GeV$^2$. Likewise for temperature variation, same set of $eB$ values are 
chosen for two different plots in the third row with fixed 
$\mu_N$(200 MeV).  }
\label{massv}
%The case with zero $eB$ is also plotted for comparison.  }
\end{figure}

 To analyze the markedly different behaviour  of the thermal contribution in the  effective masses of  $\pi^+$ and $\pi^-$, the real part of the self-energy is shown in 
 Fig.\ref{reprt}. Real part of the thermal contribution is plotted as a function of $q^0$, keeping the magnitude of the  $z$-component of the external momentum fixed at
 200 MeV. With $eB$=0.1 GeV$^2$ and T=160 MeV, the oscillatory behaviour in case of  charged pions can be observed for $\mu_N$ =0 as well as for $\mu_N$ =200 MeV.
 At vanishing $\mu_N$,   
 the self-energies of $\pi^+$ and $\pi^-$ are almost superimposed. The slight difference is present  because of the non-vanishing $q_z$. However, the introduction 
 of large $\mu_N$  reduces the  oscillations for $\pi^-$ only in the positive values of $q^0$ while they get enhanced in the domain of  negative $q^0$. Exactly the 
 opposite behaviour is seen  in case of $\pi^+$. This follows from the fact that $\Pi^+_\eta(-q,\mu_N,T)=\Pi^-_\eta(q,\mu_N,T)$ 
 as mentioned earlier in  Eq.(\ref{pi+pi-}).
 %Now, 
 %for the calculation of the effective mass,  we are only interested  in the kinematic domain  of the real pions.
 Thus the difference in the behaviour 
 of the effective mass in case of charged pions is attributed to the asymmetric behaviour of the real part of the thermal self-energy in presence of non-zero chemical 
 potential. 
 Fig.\ref{massv} describes the external parameter dependencies of the effective mass of pions. At a given temperature T=160 MeV,  variation of effective mass as a function 
 of $eB$ is shown in Fig.4(a)  and 4(b) which correspond to two different values of chemical potentials, $\mu_N$ =50 MeV and 200 MeV respectively. Clearly, the 
 effective mass shows non-trivial oscillatory behaviour in case of charged pions  as expected from earlier discussions. Moreover, one should notice that with the increase 
 of $\mu_N$, the charged pions behave differently. The oscillations in effective mass of $\pi^-$ gets reduced with the  increase in $\mu_N$ while it  enhances 
 the oscillations of $\pi^+$. 
 It is also worth mentioning that apart from the anticipated mass splitting,  there exist small windows within the given range of the magnetic field values,  
 in which the  mass hierarchy of different pion species gets altered. Moreover, the possibility of a large window exists due to the fact that, for $eB>0.1$ GeV$^2$ the  
 decreasing rate of the effective mass for $\pi^+$ is higher in comparison with that of $\pi^-$ i.e  at even higher values of magnetic field, the negatively charged 
 pions become more massive than positively charged pions. However, in extremely large 
 magnetic background, the chiral power counting does  not hold anymore \cite{colucci} which imposes serious restrictions on the validity of the model calculation.
  
 Effective mass as a function of chemical potential is presented  in Fig.4(c) and 4(d) for two different values of magnetic field, $eB$=0.1 GeV$^2$ and 
 0.2 GeV$^2$ keeping the temperature fixed at T=160 MeV.  In a similar fashion, the temperature dependence is plotted in Fig.4(e) and 4(f) with constant $\mu_N$=200 MeV 
 and with 
  $eB$=0.1 GeV$^2$ and  0.2 GeV$^2$ as before. In case of $\pi^+$ and $\pi^0$,  the effective mass slightly increases with T whereas for $\pi^-$, it almost remain 
  independent of
  temperature variation which is analogous to the $\mu_N$ dependence. However, unlike the temperature dependence, careful observation suggests that for $eB=0.2$ GeV$^2$, 
  $m^\ast _{\pi^-}$ follows a decreasing 
  trend with $\mu_N$.  Moreover,  in  plots with higher $eB$ value, there exists a noticeable  initial mass difference between neutral and charged pions. One can also 
  observe that the rate of increase of effective mass as a function of T as well as $\mu_N$ for all the pion species gets reduced for higher magnetic fields. 
 %\newpage 
%\begin{center}
%%%%%%%%%%%%%%%%%%%%%%%%%%%%%%%%%%%%%%%%%%%%%%%%%%%%%%%%%%%%%%%%%%%%%%%%%%%%%%%%%%%%%%%%%%%%%%%%%%%%%%%%%%%%%%%%%%%%%%%%%%%%%%%%%%%%%%%%%%%%%%%%%%%%%%%%%%%%%%%%%%%
 
\section{summary and conclusions}
We have evaluated the one loop pion self-energy in presence of constant homogeneous magnetic field for finite temperature and  chemical potential. As far as 
the strength of the external magnetic field 
is concerned, we have not made any approximation and  
used the complete form of the fermionic propagator represented in terms of  a sum over infinite landau levels. 
We have used the real time formalism in the evaluation of the thermal part of pion self-energy. We have solved LL dependent DSE to obtain the effective masses as a 
function of different external parameters. It is shown that by taking the trivial Landau shift  term, the effective mass increases with $eB$ for the charged pions.
Although the real part of the self-energy depends on $T$, $\mu$ and $eB$ it is sub-leading in comparison to the trivial Landau shift. Thus the effective mass of the 
charged pions remain constant as a function of both $T$ and $\mu$ for a given $eB$. To extract the contribution of the real part of the self-energy, we also solve the 
DSE by neglecting the trivial shift. It is  shown that the  effective masses of the charged pions possess oscillatory behaviour. 
However, the same oscillatory behaviour is not seen in 
case of the neutral pions. We have also shown that the oscillatory behaviour with finite chemical potential is not similar for $\pi^+$ and $\pi^-$. With increasing 
chemical potential, the oscillation in the effective mass of positive pion  is found to be enhanced while that of $\pi^-$ gets reduced. Along  with the thermal contribution,
the magnetic field dependent vacuum contribution is also taken into account. Our results suggest that the external magnetic field dependent vacuum part 
of the self-energy  significantly influences  not only  the quantitative behaviour  but also  the qualitative behaviour of the effective mass.

\section{acknowledgement}
SG acknowledges Centre for Nuclear Theory, VECC for support.

\begin{comment}

\begin{figure}
 \subfloat[]{\includegraphics[width=2.5in,angle=-90]{d1.eps}}
 \subfloat[]{\includegraphics[width=2.5in,angle=-90]{d2.eps}}
 \caption{Real parts of the thermal contributions are plotted as a function of $q^0$ for two different chemical potential, $\mu_N$=0 and 200 MeV with 
 $|q_z|$=200 MeV and external parameters $eB=0.1$ GeV$^2$ and T=160 MeV}
\end{figure}
\end{comment}

\appendix
\numberwithin{equation}{section}
\section{Evaluation of the summation}
In Eq.(\ref{sum}) we find that the vacuum contribution for the charged pion self-energy possess the  sum of three different infinite series $\mathcal{S}_1$ , $\mathcal{S}_2$
and $\mathcal{S}_3$. Here we discuss the procedure to obtain the compact expressions for these summations one by one. These compact expressions will
be  useful for evaluating the subsequent $\kpr$ integral as will be seen below.
\bea
\mathcal{S}_1&=&\sum_n^\infty(-1)^nL_{n-1}^1(2\a_p)\frac{1}{\D_n}\nn\\
&=&\sum_n(-1)^nL_{n-1}^1(2\a_p)\int_0^1\frac{dz}{\mu}z^tz^{2nxeB/\mu}\hspace{0.5cm}\mbox{with\,\,}t=\frac{\D}{\mu}-(1-x)\frac{\kpr^2}{\mu}-1\nn\\
&=&\int_0^1\frac{dz}{\mu}z^t(-z^{2xeB/\mu})(1+z^{2xeB/\mu})^{-2}\exp\Big[\frac{2\a_p(-z^{2xeB/\mu})}{-z^{2xeB/\mu}-1}\Big]
\eea
where we have used the identity
\bea
\sum_{n=0}^\infty L_n^\a(x)z^n=(1-z)^{-\a-1}\exp\Big[\frac{xz}{z-1}\Big]\hspace{0.5cm}\mbox{for\,\,\,}|z|\le1.
\eea
In a similar way with $\theta=z^{2xeB/\mu}$ we find 
\bea
\mathcal{S}_2&=&\int_0^1\frac{dz}{\mu}z^t\exp\Big(\frac{2\a_p\theta}{1+\theta}\Big)\Big[B_0-\frac{4xeBq^2\a_p\theta}{1+\theta}\Big]\hspace{0.5cm}\mbox{where}\\
\sum_{n=0}^\infty n\Big[L_n(x)-L_{n-1}(x)\Big]z^n&=&\frac{-xz}{(1-z)^2}\exp\Big[\frac{-xz}{1-z}\Big]\hspace{0.5cm}\mbox{for\,\,\,}|z|\le1
\eea
is used. $\mathcal{S}_3$ is a divergent series and to extract the momentum dependent finite part we use derivative regularization as follows.
\bea 
\mathcal{S}_3&=&\sum_n^\infty(-1)^n\big(L_n(2\a_p)-L_{n-1}(2\a_p)\big)\ln\frac{\D_n}{\mu_0}\nn\\
\frac{\partial\mathcal{S}_3}{\partial\qp^2}&=&\sum_n^\infty (-1)^n(L_n(2\a_p)-L_{n-1}(2\a_p))\frac{(-x)(1-x)}{\D_n}\nn\\
&=&\int_0^1\frac{dz}{\mu}z^t x(x-1)\exp\Big(\frac{2\a_p\theta}{1+\theta}\Big)\nn\\
\mathcal{S}_3&=&\int_0^1\frac{dz}{\ln z}z^t\exp\Big(\frac{2\a_p\theta}{1+\theta}\Big)\hspace{0.3cm}\mbox{where we use}\nn\\
\sum_{n=0}^\infty\Big[L_n(x)-L_{n-1}(x)\Big]z^n&=&\exp\Big[\frac{-xz}{1-z}\Big]\hspace{0.5cm}\mbox{for\,\,\,}|z|\le1.
\eea
It might seem that the scale is absent here but in fact is hidden in $\theta=z^{2xeB/\mu}$. Moreover, one can observe  that the $\mathcal{S}_3$ is obtained 
after an indefinite integral over $\qp^2$ which must contain an integration constant independent of $\qp^2$. In fact this constant must be infinity as the series we
started with is  divergent in nature. However this procedure extracts out the finite momentum dependent part that we require and
the infinite contribution can be taken care  by redefining the scale $\mu$ in such a way that it renormalizes the bare mass to the physical one. 
The vacuum self-energy now becomes
\bea 
(\Pi_+)_{\mbox{vac}}&=&g^2_{\pnn}\frac{2}{\pi}\int_0^1dx\int\frac{d^2\kpr}{(2\pi)^2}e^{-\a_p}\int_0^1\frac{dz}{\mu}
e^{t\ln z}\exp\Big(\frac{2\a_p\theta}{1+\theta}\Big)
\Big[-\frac{4A\theta}{(1+\theta)^2}+B_0-\frac{4xeBq^2\a_p\theta}{(1+\theta)^2}-\frac{\mu}{\ln z}\qpr^2\Big]\,\,\,\hspace{0.3cm}\nn\\
&=&g^2_{\pnn}\frac{2}{\pi}\int_0^1dx\int_0^1\frac{dz}{\mu}\int\frac{d^2\kpr}{(2\pi)^2}z^{\frac{\D}{\mu}-1}\exp\Big[-\eta(\kpr+y\qpr)^2\Big]z^{-y(1-x)\qpr^2/\mu}\nn\\
&\times&\Big[-\frac{4A\theta}{(1+\theta)^2}+B_0-\frac{4xeBq^2\a_p\theta}{(1+\theta)^2}-\frac{\mu}{\ln z}\qpr^2\Big]\hspace{0.5cm}\mbox{where}\nn\\
\eta&=&(1-x)\frac{\ln z}{\mu}+\frac{1}{eB}\mbox{tanh}(x\frac{eB}{\mu}\ln z)\nn\\
y&=&\frac{1}{\eta eB}\mbox{tanh}(x\frac{eB}{\mu}\ln z).
\eea
Now, we shift the $\kpr\rightarrow\kpr-y\qpr$. Droping the odd terms we get 
\bea 
(\Pi_+)_{\mbox{vac}}&=&g^2_{\pnn}\frac{2}{\pi}\int_0^1dx\int_0^1\frac{dz}{\mu}z^{\frac{\D}{\mu}-y(1-x)\frac{\qpr^2}{\mu}-1}\int\frac{d^2\kpr}{(2\pi)^2}e^{-\eta\kpr^2}\nn\\
&\times&\Big[-\frac{4\theta}{(1+\theta)^2}\widetilde{A}+\widetilde{B}_0+\frac{4\theta}{(1+\theta)^2}xq^2\{\kpr^2+(1-y)^2\qpr^2\}-\frac{\mu}{\ln z}\qpr^2\Big]\hspace{0.5cm}
\mbox{with}\nn\\
\widetilde{A}&=&q^2(\kpr^2+y^2\qpr^2)+2x\qp^2\qpr^2-y\qpr^2\{(1+2x)\qp^2-\qpr^2\}-2\{(\qpr\cdot\kpr)^2+y^2\qpr^4\}\hspace{0.5cm}\mbox{and}\nn\\
\widetilde{B_0}&=&-2m^2q^2+2(x-y)(1-x)\qp^2\qpr^2+(1-x)q^2(\kpr^2+y^2\qpr^2).
\eea
Now, the 2 dimensional $\kpr$ integral  can be easily evaluated using the standard Gaussian integral identities given by
\bea
\int d^2\kpr e^{-\eta\kpr^2}&=&-\frac{\pi}{\eta} \nn\\
\int d^2\kpr \kpr^2 e^{-\eta\kpr^2}&=&-\frac{\pi}{\eta^2}\nn\\
\int d^2\kpr (\qpr\cdot\kpr)^2e^{-\eta\kpr^2}&=&-\frac{\pi}{2\eta^2}\qpr^2\hspace{0.5cm}\mbox{all with\,\,}\mathcal{R}e(\eta)<0.
\eea


\newpage
 
 
\begin{thebibliography}{99}
\bibitem{ln871}D.~Kharzeev, K.~Landsteiner, A.~Schmitt and Ho-Ung Yee, Lec.Notes in Phys \textbf{871}. 
\bibitem{prep442}J.~M.~Lattimer and M.~Prakash,  Phys. Rep.~\textbf{442}, 109(2007).
\bibitem{prl95}E.~J Ferrer, V.~de la Incera and C.~Manuel, Phys. Rev. Lett. \textbf{95}, 152002 (2005); Nucl. Phys. \textbf{B747}, 88 (2006).
\bibitem{prd76}E.~J Ferrer and  V.~de la Incera, Phys. Rev.~\textbf{D 76}, 045011 (2007).
\bibitem{prl100}K.~Fukushima and H.~J.~Warringa, Phys. Rev. Lett. \textbf{100}, 032007 (2008).
\bibitem{prdN76}J.~L.~Noronha and I.~A.~Shovkovy Phys. Rev.~\textbf{D 76}, 105030 (2007).
\bibitem{prl105}B.~Feng, D.-F.~Hou, H.-C.~Ren and P.-P. Wu,   Phys. Rev. Lett. \textbf{105}, 042001 (2010) 
\bibitem{prd82}S.~Fayazbakhsh and N.~Sadooghi,  Phys. Rev.~\textbf{D 82}, 045010 (2010); \textbf{83}, 025026 (2011).
\bibitem{ijmp}V.~Skokov, A.~Y.~Illarionov and V.~Toneev, Int. J. Mod. Phys.~\textbf{A 24}, 5925(2009)
\bibitem{plb265}T.~Vachaspati, Phys. Lett.~\textbf{B 265}, 258(1991).
\bibitem{prdW86}K.~L.~Wang, S.~X.~Qin, Y.~X.~Liu, L.~Chang, C.~D.~Roberts and S.~M.~Schmidt, Phys. Rev.~\textbf{D 86}, 114001(2012).


\bibitem{prd82M}M. D'Elia, S.~Mukherjee and F.~Sanfilippo, Phys. Rev.~\textbf{D 82}, 051501 (2010).
\bibitem{prd83}M. D'Elia and F.~Negro, Phys. Rev.~\textbf{D 83}, 114028 (2011).

\bibitem{tmph90} K.~G. Klimenko, Theor. Math. Phys.~\textbf{90}, 1 (1992). 
\bibitem{prl73}V. P. Gusynin, V. A. Miransky, and I. A. Shovkovy, Phys. Rev. Lett.~\textbf{73} (1994) 3499;Erratum  Phys. Rev. Lett. 76, 1005  
\bibitem{nuclphyb563}V.P. Gusynin, V.A. Miransky, I.A. Shovkovy, Nucl. Phys. \textbf{B 563}, 361 (1999).
\bibitem{prd54}C. N. Leung, Y. J. Ng, and A. W. Ackley, Phys. Rev.~\textbf{D 54}, 4181 (1996).
\bibitem{prd68}E. Elizalde, E. J. Ferrer, and V. de la Incera, Phys. Rev.~\textbf{D 68}, 096004 (2003).
\bibitem{prd89}E. J. Ferrer, V. de la Incera, I. Portillo, and M. Quiroz, Phys. Rev.~\textbf{D 89}, 085034 (2014).
\bibitem{prd81}J. K. Boomsma and D. Boer, Phys. Rev.~\textbf{D 81}, 074005 (2010).
\bibitem{prd84}B. Chatterjee, H. Mishra and A. Mishra, Phys. Rev.~\textbf{D 84}, 014016 (2011).
\bibitem{prc83}S. S. Avancini, D. P. Menezes and C. Providencia, Phys. Rev.~\textbf{C 83}, 065805 (2011).
\bibitem{prd83M}M. Frasca and M. Ruggieri, Phys. Rev.~\textbf{D 83}, 094024 (2011).
\bibitem{prc83A}A. Rabhi and C. Providencia, Phys. Rev.~\textbf{C 83}, 055801 (2011).
\bibitem{prd82R}R. Gattto and M. Ruggieri, Phys. Rev.~\textbf{D 82}, 054027 (2010); ~\textbf{D 83}, 034016 (2011).
\bibitem{prd83K}K. Kashiwa, Phys. Rev.~\textbf{D 83},, 117901 (2011).
\bibitem{prd85}J. O. Andersen and R. Khan, Phys. Rev.~\textbf{D 85}, 065026 (2012).
\bibitem{jhep08}J. O. Andersen and A. Tranberg, J. High Energy Phys. \textbf{08}  002 (2012).
\bibitem{prd82A}A. J. Mizhar, M. N. Chernodub and E. S. Fraga, Phys. Rev.~\textbf{D 82}, 105016 (2010).
\bibitem{prd85S}V. Skokov, Phys. Rev.~\textbf{D 85}, 034026 (2012).
\bibitem{prd84D}D. C. Duarte, R. L. S. Farias and R. O. Ramos, Phys. Rev.~\textbf{D 84}, 083525 (2011).
\bibitem{prd86E}E. S. Fraga and L. F. Palhares, Phys. Rev.~\textbf{D 86} 016008 (2012).

\bibitem{jhep1202}G. S. Bali, F. Bruckmann, G. Endrodi, Z. Fodor, S. D. Katz, S. Krieg, A. Schafer, K. K. Szabo, J. High Energy Phys. \textbf{02} 044 (2012).
\bibitem{prdB86}G. S. Bali, F. Bruckmann, G. Endrodi, Z. Fodor, S. D. Katz and  A. Schafer, Phys.Rev.~\textbf{D 86}, 071502 (2012).
\bibitem{jhep04}G. S. Bali, F. Bruckmann, G. Endrodi, F. Gruber, A. Schaefer, J. High Energy Phys. \textbf{04} 130 (2013).
\bibitem{prd90}V.~ G. Bornyakov, P. V. Buividovich, N. Cundy, O. A. Kochetkov, and A. Schäfer, Phys. Rev.~\textbf{D 90}, 034501 (2014).
\bibitem{prc90}R. L. S. Farias, K. P. Gomes, G. Krein, and M. B. Pinto, Phys. Rev.~\textbf{C 90}, 025203 (2014).
\bibitem{prd91}E. J. Ferrer, V. de la Incera, and X. J. Wen, Phys. Rev. \textbf{D 91}, 054006 (2015). 
\bibitem{plb759}A. Ayala, C.A. Dominguez, L.A. Hernández, M. Loewe, R. Zamora, Phys. Lett. \textbf{B 759}, 99 (2016).
\bibitem{jhep1304}F.~Bruckmann, G.~Endrodi, T.~G.~Kovacs, J. High Energy Phys. \textbf{04} 112 (2013).

\bibitem{rmp88}J.~O.~Andersen and W.~R.~Naylor, Rev.~Mod.~Phys. \textbf{88}, 025001.
\bibitem{plb758}S.~Mao, Phys. Lett. \textbf{B 758}, 195 (2016).

\bibitem{prd79}E.~S.~Fraga, L.~F.~Palhares and  C.~Villavicencio, Phys. Rev.~\textbf{D 79}, 014021 (2009).
\bibitem{prd86}J.~O.Andersen,  Phys. Rev.~\textbf{D 86}, 025020 (2012).
\bibitem{epjc76}R.~Zhang, W.~Fu and Y.~Liu, Eur. Phys. J. \textbf{C 76}, 307 (2016).



\bibitem{colucci}G.~Colucci, E.~S.~Fraga, A.~Sedrakian, Phys. Lett.~\textbf{B 728}, 19(2014).
\bibitem{aguirre}R.~M.~Aguirre, arxiv:1612.00327v1 [nucl-th] (2016).

\bibitem{spa}S. P. Adhya, M. Mandal, S. Biswas, P. K. Roy, Phys. Rev. \textbf{D 93}, 074033 (2016). 
\bibitem{me}S. Ghosh, A. Mukherjee, M. Mandal, S. Sarkar, P. Roy, Phys. Rev. \textbf{D 94}, 094043 (2016). 

\bibitem{prc62}O. Krehl, C. Hanhart, S. Krewald and J. Speth, Phys. Rev. \textbf{C 62}, 025207.


\bibitem{olivo}J. C. D’Olivo, José F. Nieves, and S. Sahu, Phys. Rev. \textbf{D 67}, 025018  (2003).
\bibitem{prd92}A Ayala, C. A. Dominguez, L. A. Hernandez, M. Loewe, J. C. Rojas, and C. Villavicencio, Phys. Rev. \textbf{D 92}, 016006 (2015).
\bibitem{nuphyb462}V.P. Gusynin, V.A. Miransky, I.A. Shovkovy, Nucl. Phys. \textbf{B 462} (1996) 249.
\bibitem{sirbook} S. Mallik and  S. Sarkar, {\it Hadrons at Finite Temperature}, Cambridge Monographs on Mathematical Physics.




\end{thebibliography}
 \end{document}